\documentstyle[11pt,aaspp4]{article}
\lefthead{Hogerheijde et al.}
\righthead{Envelope structure of YSOs in Serpens}
\begin{document}



\title{Envelope structure of deeply embedded young stellar objects in
the Serpens Molecular Cloud}

\author{Michiel R. Hogerheijde\altaffilmark{1},
   Ewine F. van~Dishoeck, and Jante M.~Salverda\altaffilmark{2},}
\affil{Sterrewacht Leiden, P.O. Box 9513, 2300 RA, Leiden, The
Netherlands}
\authoraddr{Sterrewacht Leiden, P.O. Box 9513, 2300 RA Leiden, The
Netherlands}
\and
\author{Geoffrey A. Blake}
\affil{Division of Geological and Planetary Sciences, California
Institute of Technology, MS 150--21, Pasadena, CA 91125} 
\altaffiltext{1}{Currently at: Astronomy Department, University of
California at Berkeley}
\altaffiltext{2}{Currently at: Department of Physics and Astronomy, 
   Vrije Universiteit, Amsterdam, The Netherlands}



\begin{abstract}

Aperture-synthesis and single-dish (sub) millimeter molecular-line and
continuum observations reveal in great detail the envelope structure
of deeply embedded young stellar objects (SMM~1 = FIRS~1, SMM~2,
SMM~3, SMM~4) in the densely star-forming Serpens Molecular
Cloud. SMM~1, 3, and 4 show partially resolved ($>2''=800$ AU)
continuum emission in the beam of the Owens Valley Millimeter Array at
$\lambda=3.4$--1.4~mm. The continuum visibilities accurately constrain
the density structure in the envelopes, which can be described by a
radial power law with slope $-2.0\pm 0.5$ on scales of 300~AU to 8000
AU. Inferred envelope masses within a radius of 8000 AU are 8.7, 3.0,
and 5.3 $M_\odot$ for SMM~1, 3, and 4, respectively. A point source
with 20\%--30\% of the total flux at 1.1~mm is required to fit the
observations on long baselines, corresponding to warm envelope
material within $\sim 100$~AU or a circumstellar disk. No continuum
emission is detected interferometrically toward SMM~2, corresponding
to an upper limit of $0.2$ $M_\odot$ assuming $T_{\rm d}=24$~K. The
lack of any compact dust emission suggests that the SMM~2 core does
not contain a central protostar.

Aperture-synthesis observations of the $^{13}$CO, C$^{18}$O, HCO$^+$,
H$^{13}$CO$^+$, HCN, H$^{13}$CN, N$_2$H$^+$ 1--0, SiO 2--1, and SO
$2_2$-$1_1$ transitions reveal compact emission toward SMM~1, 3, and
4. SMM~2 shows only a number of clumps scattered throughout the
primary field of view, supporting the conclusion that this core does
not contain a central star. The compact molecular emission around
SMM~1, 3, and 4 traces $5''$--$10''$ (2000--4000 AU) diameter cores
that correspond to the densest regions of the envelopes, as well as
material directly associated with the molecular outflow. Especially
prominent are the optically thick HCN and HCO$^+$ lines which show up
brightly along the walls of the outflow cavities.  SO and SiO trace
shocked material, where their abundances may be enhanced by 1--2
orders of magnitude over dark-cloud values.

A total of 31 molecular transitions have been observed with the James
Clerk Maxwell and Caltech Submillimeter telescopes in the 230, 345,
490, and 690~GHz atmospheric windows toward all four sources,
containing, among others, lines of CO, HCO$^+$, HCN, H$_2$CO, SiO, SO,
and their isotopomers. These lines show 20--30 km~s$^{-1}$ wide line
wings, deep and narrow (1--2 km~s$^{-1}$) self-absorption, and 2--3
km~s$^{-1}$ FWHM line cores. The presence of highly excited lines like
$^{12}$CO 4--3 and 6--5, $^{13}$CO 6--5, and several H$_2$CO
transitions indicates the presence of material with temperatures
$\gtrsim 100$ K. Monte-Carlo calculations of the molecular excitation
and line transfer show that the envelope model derived from the dust
emission can successfully reproduce the observed line intensities. The
depletion of CO in the cold gas is modest compared to values inferred
in objects like NGC~1333 IRAS~4, suggesting that the phase of large
depletions through the entire envelope is short-lived and may be
influenced by the local star-formation density. Emission in high
excitation lines of CO and H$_2$CO requires the presence of a small
amount of $\sim 100$ K material, comprising less than 1\% of the total
envelope mass and probably associated with the outflow or the
innermost region of the envelope.  The derived molecular abundances in
the warm ($T_{\rm kin}>20$ K) envelope are similar to those found
toward other class~0 YSOs like IRAS 16293$-$2422, though some species
appear enhanced toward SMM~1. Taken together, the presented
observations and analysis provide the first comprehensive view of the
physical and chemical structure of the envelopes of deeply embedded
young stellar objects in a clustered environment on scales between
1000 and 10,000 AU.

\end{abstract}

\keywords{ISM: molecules --- stars: formation ---
stars: low mass, brown dwarfs --- stars: pre--main sequence}



\section{Introduction}

The earliest stages of star formation are represented by deeply
embedded, class~0 young stellar objects (YSOs; Andr\'e, Ward-Thompson,
\& Barsony \markcite{and93}1993). These sources have spectral energy
distributions (SEDs) which are well fit by a single black body curve
of $T_{\rm eff}\lesssim 30$~K, are undetected at wavelengths shorter
than $\sim 10$ $\mu$m, and still have most of their mass in a
circumstellar envelope. Thus, class~0 sources are ideal objects to
investigate the physical and chemical conditions during the earliest
phases of the star-forming process, to test theoretical models of
cloud core collapse (e.g., Shu \markcite{shu77}1977; Terebey, Shu, \&
Cassen \markcite{ter84}1984; Fiedler \& Mouschovias
\markcite{fie92}1992; Boss \markcite{bos93}1993), and to study the
influence of the bipolar outflow on the structure and chemistry of the
envelope.  This paper presents single-dish and aperture synthesis
observations of dust and molecular lines of four class~0 candidates in
the Serpens Molecular Cloud, tracing the structure and chemistry of
their envelopes on 1000--10,000 AU scales ($3''$--$30''$). These data
have higher spatial resolution than previous studies, and show a
detailed picture of the structure and chemistry of both the inner and
outer regions of the envelopes.

Previous dust continuum and molecular line observations of the
envelopes of class~0 YSOs used mostly single-dish telescopes (e.g.,
NGC 1333 IRAS 4A and 4B, Blake et~al.\ \markcite{gab95}1995; IRAS
16293$-$2422, Walker, Carlstrom, \& Bieging \markcite{wal93}1993;
Blake et~al.\ \markcite{gab94}1994, van Dishoeck et~al.\
\markcite{evd95}1995). These observations show that the outer
envelopes are dense, with $n_{\rm H_2}\approx 10^6$--$10^7$ cm$^{-3}$,
and cold, at $T_{\rm d} \approx 30$ K, and that the molecular
abundances in some objects may be significantly depleted by freezing
out onto dust grains (e.g, Blake et~al.\ \markcite{gab95}1995; see
also Mundy \& McMullin \markcite{lgm97}1997). Surveys in HCO$^+$
(Gregersen et~al.\ \markcite{gre97}1997), and in H$_2$CO, CS, and
N$_2$H$^+$ (Mardones et~al.\ \markcite{mar97}1997) indicate that
line-profile asymmetries predicted for infalling envelopes are more
readily observed during the deeply embedded class~0 phase than toward
more evolved class~I objects.  Also, class~0 YSOs are often found to
drive highly collimated outflows (e.g., Andr\'e et~al.\
\markcite{and90}1990; Guilloteau et~al.\ \markcite{gui92}1992; Zhang
et~al.\ \markcite{zha95}1995; Blake et~al.\ \markcite{gab95}1995;
Gueth et~al.\ \markcite{gue97}1997). These outflows affect the
structure of the molecular envelope, as well as its chemistry. In
material heated by the outflow to 60~K or more, molecules are released
from the grain surfaces into the gas phase, and shocks can destroy
dust particles (e.g., van~Dishoeck et~al.\ \markcite{evd95}1995; Blake
et~al.\ \markcite{gab95}1995; Bachiller \& P\'erez Guti\'errez
\markcite{bac97}1997). Several important questions about the envelopes
around class~0 sources are unanswered. How does their density
structure compare to theoretical models of protostellar collapse?  Is
the structure different in clustered regions compared with isolated
objects? To what extent is their structure and chemistry influenced by
the outflow?  How strongly are molecular abundances depleted in the
cold, dense regions?

The Serpens Molecular Cloud provides a particularly good opportunity
to study several deeply embedded class~0 objects originating from the
same molecular cloud. Casali, Eiroa, \& Duncan \markcite{cas93}(1993)
have detected four submillimeter continuum sources without any
near-infrared counterparts (SMM~1, 2, 3, and 4). Most previous studies
of these objects have been performed with single-dish telescopes at
$\sim 15''$ resolution (e.g., White et~al.\ \markcite{whi95}1995; Hurt
et~al.\ \markcite{hur96b}1996b). This paper presents single-dish and
aperture-synthesis observations of molecular lines and dust continuum
at (sub) millimeter wavelengths toward SMM~1, 2, 3, and 4, with
spatial resolutions between $1''$ and $20''$ ($\sim 400$--$8000$
AU). The protostellar nature of three sources, SMM~1, 3, and 4, is
confirmed on the basis of the interferometer results, while SMM~2
appears to be a warm cloud condensation without any central
source. The continuum observations allow a determination of the
density and temperature structure of the envelopes surrounding SMM~1,
3, and 4. The molecular line data can be explained by this same
envelope model, but require that $\lesssim 1\%$ of the gas has a much
higher temperature of $\sim 100$ K. The aperture-synthesis
observations trace $5''$--$10''$ (2000--4000 AU) cores around the
YSOs, as well as the interaction of the outflow with surrounding
material. This study of a group of deeply embedded YSOs has the
spatial resolution required to directly sample the protostellar
envelopes in a clustered environment. The Serpens results will be
compared to class~0 objects in other clouds, as well as to a set of
more evolved class~I sources studied in emission of the same molecular
transitions by Hogerheijde et~al.\ (\markcite{mrh97a}1997,
\markcite{mrh98a}1998).

The outline of the paper is as follows. Section 2 introduces the
characteristics of the Serpens Molecular Cloud and its embedded
YSOs. After presenting the observations in \S 3, we discuss the
results of the continuum measurements in \S 4.1, and construct an
envelope model in \S 4.2.  Section 5 analyzes in detail the molecular
line emission observed in the interferometer (\S 5.1) and single-dish
beams (\S 5.2) toward the individual sources. In \S 5.3 the molecular
line emission is compared to model predictions based on the envelope
structure derived from the dust continuum. The results are further
discussed and compared to other class~0 and class~I YSOs in \S 6.
Finally, \S 7 summarizes the main findings of the paper.


\section{The Serpens Molecular Cloud and its embedded protostars}

The Serpens Molecular Cloud appears to be forming a loosely bound
cluster of low- to intermediate-mass stars. Eiroa \& Casali
\markcite{eir92}(1992) have identified 51 near-infrared sources as
T~Tauri stars (see also Giovannetti et~al.\ \markcite{gio98}1998).
Nordh et~al.\ \markcite{nor82}(1982) and Harvey et~al.\
\markcite{har84}(1984) first identified SMM~1, also called FIRS~1, as
a deeply embedded YSO. Subsequent submillimeter continuum maps by
Casali et~al.\ \markcite{cas93}(1993) resulted in the detection of
four strong sources without near-infrared counterparts, SMM~1--4.
Recently, Testi \& Sargent \markcite{tes98}(1998) carried out an
extensive survey of the Serpens cloud with the Owens Valley Millimeter
Array in CS 2--1 and 3~mm continuum, identifying 32 cores with masses
in the range of 0.4--16 $M_\odot$. The SMM 1--4 sources stand out as
the brightest of these cores, and represent a second, more recent
phase of star formation in the Serpens cloud compared to the
near-infrared sources. Their spectral energy distributions are well
fit by single blackbodies with temperatures of 20--27 K,
characteristic of class~0 YSOs (Hurt \& Barsony
\markcite{hur96a}1996a). Observations of $^{12}$CO, HCO$^+$, and
H$_2$CO lines indicate the presence of warm ($T_{\rm kin}=40$--190 K)
and dense ($n_{\rm H_2} \approx 2\times10^6$ cm$^{-3}$) gas associated
with bipolar outflows, as evidenced by broad line wings (White,
Casali, \& Eiroa \markcite{whi95}1995; Hurt, Barsony, \& Wootten
\markcite{hur96b}1996b). 

Near-infrared observations also reveal evidence for outflows. Herbst,
Beckwith, \& Robberto \markcite{her97a}(1997) detected a number of
H$_2$ knots, possibly a jet, emanating from SMM~3, whereas Eiroa
et~al.\ \markcite{eir97}(1997) report the detection of a similar
string of H$_2$ knots possibly associated with SMM~4.  SMM1 shows
several signs of energetic activity, such as H$_2$O maser emission
(Dinger \& Dickinson \markcite{din80}1980; Rodr\'{\i}guez et~al.\
\markcite{rod80}1980) and near-infrared H$_2$ emission (Eiroa \&
Casali \markcite{eir89}1989). At cm wavelengths, the source is triple,
with two diametrically opposed lobes moving away at $\sim 200$
km~s$^{-1}$ from the central source.  McMullin et~al.\
\markcite{mcm94}(1994) performed an interferometric and single-dish
study of the northwestern part of the Serpens Molecular Cloud
including SMM~1 and the condensation S68~N, $2'$ north of SMM~1, using
lines of CS, CH$_3$OH, and other molecules. They find that most of the
70 $M_\odot$ in this region is distributed on extended scales,
indicative of a young age of the sources, with $\sim 10$ $M_\odot$ in
a circumstellar component around SMM~1. They infer abundances for CO
and HCO$^+$ corresponding to those found in dark clouds, while CS,
HCN, and H$_2$CO may be mildly depleted. Evidence for freezing out of
molecules on grains is also provided by direct observations of solid
CO in the Serpens Molecular Cloud by Chiar et~al.\
\markcite{chi94}(1994).

The distance to the Serpens cloud core has been the subject of much
debate. De Lara, Chavarria-K, \& L\'opez-Molina \markcite{lar91}(1991)
derive a value of $310\pm 40$ pc, based on extinction measurement of
five stars. Chiar \markcite{chi96}(1996) finds $425\pm 45$ pc based on
seven stars, including new observations of the five stars used by de
Lara et~al., all of which indicate a larger distance. One of these
stars, R16, has a derived distance of 628 pc, but is included by Chiar
because its image on the POSS plate suggests it is an embedded object
associated with the Serpens cloud. Excluding this star lowers the
distance estimate to 390 pc. Here we will adopt 400 pc as a fiducial
estimate of the distance to Serpens. Bolometric luminosities and other
quantities taken from the literature are scaled to the adopted
distance.


\section{Observations}

Table~1 lists the coordinates of SMM~1, 2, 3, and 4, together with
their bolometric luminosity scaled to a distance of 400~pc, continuum
flux at $\lambda=1.1$ mm, and estimates of the stellar mass.  The
coordinates are derived from the interferometric continuum emission,
and differ by up to $5''$ from those quoted by Casali et~al.\
\markcite{cas93}(1993) for SMM~3 and 4. Lower limits to the stellar
masses follow from the assumption that all luminosity is due to
accretion at a rate of $10^{-5}$ $M_\odot$~yr$^{-1}$, while upper
limits are the stellar masses which produce the same luminosity on the
zero-age main sequence. Table 2 gives an overview of the data
presented in this paper. The following subsections present the details
of the interferometer and single-dish observations.

\subsection{Millimeter-interferometer observations}

Observations of the transitions listed in Table 2 were obtained with the
six-element Owens Valley Radio Observatory (OVRO) Millimeter Array%
\footnote{The Owens Valley Millimeter Array is operated by the
California Institute of Technology under funding from the U.S. National
Science Foundation (\#AST96--13717).} 
between 1994 and 1997, simultaneously with the continuum emission over
a 1~GHz bandwidth at $\lambda=3.4$, 3.2, 2.7, and 1.4~mm. Data taken
in the low-resolution and equatorial configurations were combined,
resulting in a $u$-$v$ coverage with spacings between 3 and
60--80~k$\lambda$ at 3.4--2.7~mm for the observed lines and continuum,
and between 10 and 100--180~k$\lambda$ at 1.4~mm in continuum only.
This corresponds to naturally weighted, synthesized beams of
$3''$--$5''$ and $1''$ FWHM, respectively. Spectral line data were
recorded in two 64-channel bands with respective widths of 2 and
8~MHz, resulting in velocity resolutions of $\sim 0.1$ and $\sim
0.4$~${\rm km\,s^{-1}}$. The visibility data were calibrated using the
MMA package, developed specifically for OVRO (Scoville et~al.\
\markcite{sco93}1993).  The quasar PKS~1749+096 served as phase
calibrator; the amplitudes were calibrated on 3C~454.3, 3C~273, or
Neptune. The correlator passbands are calibrated using noise tube
integrations and observations of 3C~454.3 and 3C~273.

The interferometer data were edited in the usual manner by flagging a
small number of data points with clearly deviating amplitudes and
phases. The quality of the continuum data at 3.4--2.7 mm allowed
self-calibration of the visibility phases, which was subsequently
applied to the line data, decreasing the noise level by a small
amount.  Natural weighting was used to clean the data. The
interferometer dirty beam has strong N--S side lobes because of
the $+1^\circ$ declination of Serpens. In many cases the
position of the initial clean-components had to be constrained to a
$\sim 20''$ box around the source position to ensure proper
deconvolution. An additional complication for cleaning of the
molecular-line data toward SMM~3 was the presence of strong emission
from SMM~4 approximately one primary beam ($\sim 70''$) to the
south. Care had to be taken to prevent any emission from SMM~4 appearing
within the primary field of view around SMM~3. 

The reduced continuum data have RMS noise levels of 2--6
mJy~beam$^{-1}$ at 3.4--2.7 mm and 30 mJy~beam$^{-1}$ at 1.4 mm. For
the molecular line data the typical noise is $0.1$--$0.2$
Jy~beam$^{-1}$ per 125~kHz channel. Reduction and analysis of the
visibility data were carried out within the MIRIAD software package.

\subsection{Single-dish observations}

The single-dish line observations between 219 and 690 GHz were
obtained between 1995 March and 1996 August with the James Clerk
Maxwell Telescope (JCMT)%
\footnote{The James Clerk Maxwell Telescope is operated by the
Joint Astronomy Centre, on behalf of the Particle Physics and
Astronomy Research Council of the United Kingdom, the Netherlands
Organization for Scientific Research and the National Research Council
of Canada.}
and the Caltech Submillimeter Observatory (CSO)%
\footnote{The Caltech Submillimeter Observatory is operated by the
California Institute of Technology under funding from the U.S.\
National Science Foundation (\#AST96--15025).}
. The observed transitions are listed in Table 2. The single-dish
observations were reduced and analyzed with the CLASS software
package.

The JCMT observations at 230, 345, and 490~GHz
were obtained with FWHM beam sizes of $19''$, $14''$, and $10''$,
respectively. The observations were made using a position-switch of
typically $15'$--$60'$, ensuring emission-free offset positions.
Pointing accuracy is estimated to be $\sim 5''$.  As indicated in 
Table 2, maps covering regions between $20''\times 20''$ and $40''\times 
40''$ around the sources were taken in a number of lines, sampled
at $1\over 2$--$2\over 3$ beam sizes. In HCO$^+$ 3--2, fully sampled
maps were obtained {\it on-the-fly\/}. The spectra were recorded with
the Digital Autocorrelation Spectrometer, with typical velocity
resolutions of 0.05--0.1 km~s$^{-1}$. The H$_2$CO $3_{03}$--$2_{02}$
and $3_{22}$--$2_{21}$ spectra were observed in a single frequency setting
over a total bandwidth of 500~MHz and at a resolution of 0.9 km~s$^{-1}$,
also covering the HC$_3$N 24--23, C$_3$H$_2$ $5_{24}$--$4_{13}$,
CH$_3$OH $4_2$--$3_1$~E, and SO $5_5$--$4_4$ lines. The data were
converted to the main-beam antenna temperature scale using $\eta_{\rm
mb}=0.69$ (230 GHz), 0.58 (345 GHz), and 0.53 (490 GHz), obtained by
the JCMT staff from measurements of the planets. Typical RMS noise
levels are 0.1--0.3 K in 0.15 km~s$^{-1}$ wide channels.

Using the CSO, observations were obtained of $^{12}$CO and $^{13}$CO
6--5 with a FWHM beam size of $10''$, and of H$^{13}$CO$^+$ 4--3 at
$22''$, the latter of which also contains the SO $8_8$-$7_7$ line. 
The observations were made using a position switch of $15'$ for the
$^{12}$CO and $^{13}$CO spectra, and a beam switch of $180''$ for the
H$^{13}$CO$^+$ data, ensuring emission-free offset positions. Pointing
was checked regularly, and found to vary by up to $5''$. At the frequency
of the $^{12}$CO and $^{13}$CO 6--5 lines (660--690 GHz), an additional
source of positional error was the correction for the atmospheric
refraction, which is comparable to the FWHM beam size ($\sim
10''$). It is estimated that the pointing at these frequencies is no
better than $\sim 10''$. Five-point maps with $10''$ spacing were
obtained for $^{12}$CO and $^{13}$CO 6--5. The spectral lines were
recorded with the facility 50~MHz and 500~MHz bandwidth
Acousto--Optical Spectrometers (AOSs). The spectra were converted to
the main-beam antenna scale using $\eta_{\rm mb} = 0.44$
($^{12}$CO, $^{13}$CO) and $0.65$ (H$^{13}$CO$^+$, SO), obtained from
measurements of Jupiter. Resulting RMS noise levels are 0.6 K
per 0.5 km~s$^{-1}$ channel for $^{12}$CO and $^{13}$CO, and 0.1 K per
0.15 km~s$^{-1}$ for H$^{13}$CO$^+$ and SO.


\section{Dust continuum emission}

\subsection{Interferometer results}

Continuum emission at $\lambda=3.4$, 3.2, 2.7, and 1.4~mm is readily
detected by OVRO toward SMM~1, 3, and 4. Peak intensities and
integrated fluxes are listed in Table 3. No continuum emission is
detected toward SMM~2 with an upper limit of $\sim 5$ mJy at 3.4 mm.
Figure~1 shows the naturally weighted, cleaned continuum images. The
emission is mostly unresolved and symmetric about the source
position. Table~1 lists the best-fit source positions for SMM~1, 3,
and 4. Our observations, which have higher spatial resolution and
positional accuracy than do the JCMT continuum maps of Casali et~al.\
\markcite{cas93}(1993), yield positions for SMM~3 and SMM~4 nearly $5''$
west of those listed by these authors. We will adopt our best-fit
coordinates as their true positions. The best-fit
coordinates of SMM~1 agree with the radio position from Rodr\'{\i}guez
et~al.\ \markcite{rod89}(1989) to within the accuracy of the measurements.
Its emission at cm wavelengths, and the expected flat spectral index for
non-thermal radiation, indicates that $>95\%$ of the emission of SMM~1
at 3.4--1.4 mm is due to thermal emission from dust (Rodr\'{\i}guez et~al.\
\markcite{rod89}1989; McMullin et~al.\ \markcite{mcm94}1994).

Figure~2 shows the vector-averaged visibility amplitudes as functions of
projected baseline length for the observed sources and wavelengths,
averaged in 5--10 k$\lambda$ wide bins. These plots essentially give
the Fourier transform of the symmetric part of the sky brightness
about the source center. A point source has a flux independent of
baseline length, while the flux of an extended source decreases with
increasing $u$-$v$ separation. The observations show both extended and
unresolved ($<2''=800$ AU) emission towards SMM~1, 3, and 4. The
extended emission traces the envelopes surrounding the YSOs, and
the steep decrease of flux with $u$-$v$ distance suggests a radial
power-law distribution for the density. A Gaussian distribution, for
example, would appear Gaussian as well in a plot of this type. The
unresolved emission may contain contributions from the dense central
regions of the power-law envelope as well as that from a circumstellar
disk. The amplitudes on baselines $\gtrsim 60$ k$\lambda$ at the
different wavelengths indicate a spectral index of 2.0 for the
unresolved emission, consistent with optically thick, thermal emission.

\subsection{A model for the continuum emission}

The high signal-to-noise of the resolved continuum emission in the
interferometer beam, together with the single-dish (sub) millimeter
and IRAS fluxes of Casali et~al.\ \markcite{cas93}(1993) and Hurt \&
Barsony \markcite{hur96a}(1996a), provides constraints on the mass and
density distribution of the envelopes. Our modeling explicitly
includes sampling at the discrete $(u,v)$ positions of the visibility
data and the resulting resolving-out of extended emission.

Table 4 summarizes the basic parameters of the adopted envelope
model. For the density distribution we assume a radial power law,
$\rho \propto r^{-p}$, as suggested by the visibility amplitudes of
Fig. 2. The index $p$ is a free parameter of the model, and is varied
between $1$ and $3$. Theoretical models for cloud core collapse
predict slopes between $-1$ and $-2$ (e.g., Shu \markcite{shu77}1977;
Lizano \& Shu \markcite{liz89}1989).  Values for the dust emissivity
at millimeter wavelengths are taken from Ossenkopf \& Henning
\markcite{oss94}(1994), which include dust coagulation in a medium of
$10^6$ H$_2$ cm$^{-3}$ and thin ice mantles, with $\kappa_\nu (1.3\,
{\rm mm})=0.9$ cm$^{2}$~g$^{-1}$(dust) and $\kappa_\nu \propto
\nu^{1.5}$. Inner and outer radii of 100 and 8000 AU are adopted, the
exact values of which do not influence the derived parameters
significantly.

The dust temperature is approximated by a power law of index $-0.4$,
expected for a centrally heated, spherical cloud which is optically
thin to the bulk of the radiation (cf.\ Rowan-Robinson
\markcite{row80}1980; Adams \& Shu \markcite{ada87}1987). At large
radii the temperature is not allowed to drop below 8 K, corresponding
to the typical value maintained through cosmic ray heating of the
hydrogen gas. To fit the peak of the SED at 50--100 $\mu$m, dust
temperatures at 1000 AU of 27~K (SMM~1), 24~K (SMM~3), and 20~K
(SMM~4) are required, similar to the values found by Hurt \& Barsony
\markcite{hur96a}(1996a). The latter authors assume a single dust
temperature, but our derived temperature gradient is sufficiently
shallow to give similar results.  A self-consistent calculation of the
heating and cooling balance of the dust for representative envelope
parameters confirms that the temperature follows $T_{\rm d} \propto
r^{-0.4}$ outside radii of 200--400 AU (see van der Tak et~al.\
\markcite{tak98}1998 for details of the temperature
calculations). Toward smaller radii, the temperature increases more
rapidly. Since these radii are not resolved in the interferometer
observations, the associated excess emission is fitted by a simple
point source instead, with spectral index $\alpha=2.0$ (cf.\ \S
4.1).

The resulting model has three free parameters.  The total envelope
mass $M_{\rm env}$ within a radius of 8000 AU is found from the
single-dish 1.1~mm continuum fluxes observed in the $18''$ JCMT beam
(Casali et~al.\ \markcite{cas93}1993). The flux of the unresolved
point source follows from the flux on long baselines, and the density
power-law index $p$ is constrained from the change of the visibility
amplitude with $u$--$v$ separation. Figure~3 illustrates this for
models fitted to the total 1.1~mm flux with various power-law indices
and point-source fluxes. Models with a shallower density distribution
have amplitudes which decrease faster with increasing $u$-$v$
distance. The best correspondence to the observed amplitudes of SMM~1
is found for $p=2.0\pm 0.5$, which accurately fits the flux increase
toward smaller $u$-$v$ distances, and a point-source flux of 0.13 Jy
at 2.7~mm.

Similar best-fit results with $p=2.0\pm 0.5$ are found for SMM~3 and 4
with point-source fluxes at 2.7~mm of 0.05 Jy and 0.07 Jy,
respectively. The corresponding curves are drawn in Fig.~2 for all
sources. The inferred slopes of $-2.0$ for the density agree well with
theoretical predictions for very young sources (e.g., Shu
\markcite{shu77}1977). While the model curves provide very good fits
to the 3.4--2.7 mm fluxes, they do underestimate the large scale
emission at 1.4 mm. Part of this discrepancy may be explained by a
steeper frequency dependency of the dust emissivity than the $\sim
\nu^{1.5}$ of the adopted model (Ossenkopf \& Henning \markcite{oss94}1994).
It may also indicate that warm material is distributed over larger
scales than assumed in the model, in which it is confined to the inner
few hundred AU.

Corresponding envelope masses are 8.7 $M_\odot$ for SMM~1, 3.0
$M_\odot$ for SMM~3, and 5.3 $M_\odot$ for SMM~4. These values depend
on the adopted dust emissivity of Ossenkopf \& Henning
\markcite{oss94}(1994), which is uncertain by approximately a factor
of 2--3 (cf.\ also Agladze et~al.\ \markcite{agl94}1994; Pollack
et~al.\ \markcite{pol94}1994). These masses exceed the limits placed
on the stellar mass of 0.7--3.9 $M_\odot$ (SMM~1), 0.1--2.2 (SMM~3),
and 0.1--2.3 $M_\odot$ (SMM~4; Table~1), confirming the young age of
these sources and their classification as class~0 YSOs.

The point-source fluxes correspond to masses within 100~AU of 0.9,
0.4, and 0.5 $M_\odot$ for SMM~1, SMM~3, and SMM~4, respectively,
adopting a dust temperature of 100~K and optically thin
radiation. However, the self-consistent temperature calculations
suggest that the temperature exceeds 100~K by factors of a few on
these small radii, while the spectral indices of the point-source
emission indicate that the emission is optically thick at millimeter
wavelengths. These considerations make the estimated masses within
100~AU uncertain by at least a factor of a few. A small fraction of
the unresolved emission may originate in a circumstellar disk, but
observations at much higher angular resolution are required to
separate this from emission due to the inner envelope.  Using
single-baseline interferometry, Pudritz et~al.\ \markcite{pud96}(1996)
infer a flux of 0.12 Jy at $\lambda=1.4$ mm for a disk around the
class~0 YSO VLA~1623 ($d=160$ pc). Scaling to the distance of Serpens
and to $\lambda=2.7$~mm using a spectral index of 2.0, this
corresponds to only 5 mJy, suggesting that the fitted point-source
emission of 50--130 mJy toward these sources is dominated by compact
envelope material. Calvet, Hartmann, \& Strom \markcite{cal97}(1997)
infer the presence of hot dust very close to the star ($\sim 0.1$ AU)
as an explanation for the weakness of CO $v=2$--0 emission and
absorption from Class~I objects through veiling by infalling dust from
the envelope.


\section{Molecular line emission}

The dust continuum observations provide direct constraints on the
density gradient in the envelope. Molecular-line data offer a
complementary view of the density structure, and probe the influence
of the outflow on the envelope and the response of the chemistry. The
various components in the protostellar environment traced by the
observations are summarized in Table 5.

\subsection{Interferometer results}

Compact emission is detected in most transitions observed toward the
four sources. Figure~4 shows the naturally weighted, cleaned images of
the integrated intensity; Fig.~5 presents the spectra obtained within a
$5''\times 5''$ box around the image maximum, corresponding
approximately to the synthesized beam. Table~6 lists the
velocity-integrated brightness temperatures $\int T_{\rm b}dV$
averaged over $5''\times 5''$ and $20''\times 20''$ regions around the
source positions listed in Table~1. Estimates of the opacity averaged
over the line profiles follow from the observed ratios of
C$^{18}$O/$^{13}$CO, H$^{13}$CO$^+$/HCO$^+$, and H$^{13}$CN/HCN, and
the relative intensities of the hyper-fine components of HCN and
N$_2$H$^+$. Isotopic abundance ratios of [$^{13}$CO]:[C$^{18}$O]=8:1
and [HCO$^+$]:[H$^{13}$CO$^+$]=[HCN]:[H$^{13}$CN]=65:1 are assumed
(Wilson \& Rood \markcite{wil94}1994).

Extended ($\gtrsim 20''$) emission from optically thick material is
resolved out by the interferometer in $^{13}$CO, HCO$^+$, and HCN 1--0
toward all sources. This results in apparent, deep absorption features
in the spectra of Fig.~5, and negative intensities in the images of
Fig.~4. The level of resolved-out $^{13}$CO 1--0 emission toward SMM~4
is so large that no positive signal is left in the
integrated-intensity image. Using only the velocity interval of 4--8
km~s$^{-1}$, where positive emission is detected, yields a core of
$\sim 10''$ diameter around the source. Keeping in mind these high
levels of resolved-out emission, the interferometer molecular line
data can be interpreted, with caution, both qualitatively and
quantitatively (cf.\ also \S 5.3).

As discussed in \S 3.1, the pointing centers of the observations of
SMM~2, 3, and 4 are separated by only one primary-beam size, and care
had to be taken to prevent emission spilling over in the deconvolved
images. Figure~4 presents the cleaned images at the different
transitions in single panels containing all three sources. The plotted
images have been cleaned individually, since no reliable mosaic could
be obtained with the pointing centers separated by a full primary
beam. Maximum entropy deconvolution of the mosaicked images did
yield consistent results, however.

Estimates of the molecular abundances on the scales traced by the
interferometer are derived in \S 5.3 using detailed modeling of the
molecular excitation, radiative transfer, and $(u,v)$
sampling. Detailed analysis has shown that the derived abundances may
be in error by as much as a factor of 5 if the intensities listed in
Table~6 are used without going through this careful procedure
(Hogerheijde \markcite{phd98}1998; Hogerheijde \& van der Tak
\markcite{mrh98z}1998).

The following sections discuss the specific details of the
aperture-synthesis results for the individual sources. In summary, the
C$^{18}$O, $^{13}$CO, and H$^{13}$CO$^+$ lines probe $5''$--$10''$
(2000--4000 AU) cores surrounding SMM~1, 3, and 4. The optically thick
HCO$^+$ and HCN emission is associated with the walls of the outflow
cavities, while SiO and SO probably reveal shocked material where the
outflow impacts directly on the envelope. These physical components
are similar to those inferred toward other embedded YSOs (e.g.,
B5~IRS~1, Langer, Velusamy, \& Xie \markcite{lan96}1996; B1, Hirano
et~al.\ \markcite{hir97}1997; and class~I YSOs in Taurus, Hogerheijde
et~al.\ \markcite{mrh97a}1997). None of the four sources shows
detectable emission in C$^{34}$S, while in C$_3$H$_2$ only a number of
scattered clumps are detected toward SMM~1 and 4. The latter line
probably traces dense condensations in the surrounding, quiescent
cloud, and is not further considered here. The $3\sigma$ upper limit
of 3.9 K~km~s$^{-1}$ on the emission of C$^{34}$S toward SMM~1 is
consistent with the weak detection of CS 2--1 toward this source of
$\sim 6$ K~km~s$^{-1}$ by McMullin et~al.\ \markcite{mcm94}(1994).
SMM~2 does not show any compact emission, consistent with the
interpretation that this core does not contain a protostar.

\subsubsection{SMM~1}

A compact, $\sim 10''$ (4000~AU) diameter core around SMM~1 is traced
by the OVRO observations of C$^{18}$O, H$^{13}$CO$^+$, and H$^{13}$CN
1--0. Assuming local thermodynamic equilibrium (LTE), a fiducial
estimate of the kinetic temperature of 30~K, and a C$^{18}$O abundance
with respect to H$_2$ of $2\times 10^{-7}$, a mass of 0.25 $M_\odot$
is inferred over a $20''\times 20''$ region. Although this appears
much less than the mass of 8.7 $M_\odot$ inferred from the dust
emission in \S 4.2, modeling in \S 5.3 indicates that for the derived
envelope parameters only a low fraction of the mass is indeed
recovered through C$^{18}$O emission in the OVRO beam.

The emission from optically thick HCO$^+$, HCN, and $^{13}$CO lines
appears to be directly associated with the outflow. Because of the large
level of resolved-out emission in these lines, only red and blue
shifted material is traced by the interferometer, whereas the bulk of
the emission from the envelope is not seen. This is especially
apparent in $^{13}$CO, where only unresolved emission is recovered,
but with a total velocity width of almost 10 km~s$^{-1}$. The emission
of HCO$^+$ and HCN apparently traces the walls of the outflow
cavity. The position angle of the outflow is constrained by radio
measurements to roughly $-50^\circ$ (Rodr\'{\i}guez et~al.\
\markcite{rod89}1989), bisecting the two arms seen in HCN, and the
roughly cross-shaped HCO$^+$ emission. Only the peak of the emission,
centered on the protostar, has a velocity extent of $\sim 20$
km~s$^{-1}$ in the HCO$^+$ and HCN spectra. The extended `arms' lie
within $\sim 5$ km~s$^{-1}$ of the systemic velocity. H$_2$ emission
at 2.2 $\mu$m coincides with the northern HCO$^+$ and HCN arm, probably
tracing shocked material along the wall of the outflow (cf.\
Rodr\'{\i}guez et~al.\ \markcite{rod89}1989). N$_2$H$^+$ 1--0 emission
coincides with the northern outflow wall, but the lines are much
narrower than those of HCO$^+$, suggesting association with the
envelope only.

Emission from SiO 2--1 and SO $2_2$--$1_1$ coincides with the
southeast outflow lobe. The prominent blue line wing in the spectrum
at the emission peak at the southeast tip of the lobe clearly shows
the association of SiO with material in the outflow.  Using the
C$^{18}$O upper limit at the peak of the SiO and SO emission, and
assuming LTE excitation at $\sim 100$ K, yields lower limits to the
abundance of a few times $10^{-8}$ for SiO and $\sim 10^{-7}$ for
SO. Such enhanced abundances of SiO and SO are attributed to shocked
material (cf.\ Bachiller \markcite{bac96}1996; van Dishoeck \& Blake
\markcite{evd98}1998). More accurate constraints on the SiO and SO
abundances require that the missing zero-spacing flux of SiO, SO, and
C$^{18}$O be taken into account. Such detailed modeling is also
required to investigate the relation between the material traced in
SiO and SO compared to HCO$^+$ and HCN.

In $^{12}$CO 2--1 maps presented by White et~al.\
\markcite{whi95}(1995) the outflow shows a complicated structure with
the northwestern outflow lobe being mostly blue shifted, while the
southeastern lobe shows both red- and blue-shifted emission. The small
scale structure associated with the outflow also is markedly
asymmetric. In H$_2$ at 2.2 $\mu$m, and in HCO$^+$ and N$_2$H$^+$ in
the OVRO beam the northern cavity wall is most prominent, while HCN
only traces the northern and western walls.  SiO and SO emission trace
the southeastern outflow lobe. This asymmetry might be explained by a
large-scale density gradient in the ambient medium, with denser material
located to the east, coincident with the location of the bulk of the Serpens
cloud. In such a medium, the time scale to clear an outflow cavity to
the southeast may be longer than that to the northwest. SiO and SO emission
trace shock interaction of the latter outflow lobe with ambient
material, while the larger column of material is optically thick to
any associated HCO$^+$ or HCN emission. For $n_{\rm H_2}\approx 10^6$
cm$^{-3}$ and $T_{\rm kin}\approx 30$ K, a column of a few times
$10^{14}$ cm$^{-2}$ in HCO$^+$ or HCN is sufficient for the required level
of obscuration. With the envelope parameters derived in \S 4.2, this
corresponds to a density contrast of a factor of a few between both
sides of the envelope. To the northwest, HCO$^+$ and HCN emission
arises in the walls surrounding a relatively empty outflow cavity, and
H$_2$ 2 $\mu$m emission is observable where the outflow breaks out of
the cloud.

\subsubsection{SMM~2}

SMM~2 only shows emission from $^{13}$CO, HCO$^+$, H$^{13}$CO$^+$, and
HCN in clumps scattered throughout the field of view. The spectra
toward the emission peaks reveal narrow lines of 1 km~s$^{-1}$ width,
except for $^{13}$CO. This lack of central condensation is consistent
with the upper limit on the continuum emission from \S 4.1, and
supports the interpretation that the SMM~2 core does not contain a
protostar. Instead, the interferometer appears to trace irregular
structure in the extended cloud, with most of the emission distributed
on large, resolved-out scales.

\subsubsection{SMM~3}

The $^{13}$CO emission toward this source traces a $10''$ (4000 AU)
diameter core around the continuum position, with a mass of 0.08
$M_\odot$, while only weak emission is detected in C$^{18}$O. Model
calculations in \S 5.2 confirm that only a small fraction of the 3.0
$M_\odot$ envelope is recovered in $^{13}$CO emission in the OVRO
beam.  The HCO$^+$ and HCN images show elongated emission extending
over $\sim 30''$ at a position angle of $-20^\circ$. Herbst et~al.\
\markcite{her97a}(1997) have detected a string of H$_2$ emission knots
along a line with the same orientation, which they attribute to a
jet. In this interpretation, HCO$^+$ and HCN trace a highly collimated
outflow. Marginally detected SiO coincides with the south end of the
outflow. The high degree of collimation of the outflow may indicate
that SMM~3 is a particularly young YSO. Together with the projected
location of SMM~3 on the edge of the outflow of SMM~4 (see below, and
Fig.~4), this may suggest induced star formation. Barsony et~al.\
\markcite{bar98}(1998) suggest a similar scenario for the 
L1448N(A/B)$+$L1448~NW system, and for the NGC 1333 molecular cloud
core by Lefloch et~al.\ \markcite{lef98}(1998).

Although the visual extinction toward SMM~3 is still very large, it
does have the lowest inferred envelope mass of the three YSOs,
possibly because it is forming a lower mass star. It is therefore
probably less deeply embedded than SMM~1 or 4, based on the weak
detection of C$^{18}$O, the upper limits on H$^{13}$CO$^+$, and the
relatively shallow absorption features in the spectra. Weak H$^{13}$CN
emission coincides with the northern outflow lobe, indicating that HCN
may be strongly enhanced by the outflow.

\subsubsection{SMM~4}

The C$^{18}$O, H$^{13}$CO$^+$, and N$_2$H$^+$ emission traces a
$5''$--$10''$ (2000--4000 AU) core around this source with a mass of
0.15 $M_\odot$. As for SMM~1 and 3, modeling in \S 5.3 indicates that
the amount of material traced by C$^{18}$O in the OVRO beam is
consistent with the parameters of the 5.3 $M_\odot$ envelope derived
in \S 4.2. As explained above, the integrated $^{13}$CO intensity
image does not show any emission because of the large optical depth
of resolved-out extended emission, so that the $^{13}$CO spectrum
reveals a deep absorption feature close to the systemic velocity.

In HCO$^+$ and HCN the emission is again associated with the
outflow. The $^{12}$CO 2--1 single-dish observations of White
et~al. \markcite{whi95}(1995) indicate a north-south
orientation. Eiroa et~al.\ \markcite{eir97}(1997) present 2.2 $\mu$m
observations showing a possible H$_2$ jet with a position angle of
$\sim 10^\circ$ emanating from SMM~4. HCO$^+$ emission outlines the
northern outflow cavity, while HCN also traces the southern outflow
lobe. In addition, strong emission originates from a $10''$ core,
elongated perpendicular to the outflow direction. The bulk of the
HCO$^+$ and HCN emission occurs within 2--4 km~s$^{-1}$ from line
center, as seen in the spectra.  This indicates that the HCO$^+$ and
HCN lines trace envelope material which is heated, compressed, or
chemically altered by the outflow, but not entrained within the
outflow itself. In HCO$^+$ there is a 4 km~s$^{-1}$ west-to-east
velocity gradient, possibly indicating rotation in the envelope and
cavity walls. The hyperfine components in the HCN line confuse any
velocity gradient in that line. The fact that HCO$^+$ emission is
associated with the northern outflow lobe only may be connected to a
higher outflow or interaction activity on that side, in contrast to
SMM~1 where the asymmetry of the HCO$^+$ and HCN emission may be due
to increased density and opacity on the southeast side. The $^{12}$CO
2--1 outflow maps of White et~al.\ \markcite{whi95}(1995) also show
more intense outflow emission toward the north of SMM~4. A number of
marginally detected clumps of SiO are seen close to the center of
SMM~4, but no SO is found.


\subsection{Single-dish results}

This section discusses the single-dish observations. These data lack
the spatial resolution of the interferometer results, but include
higher excitation lines, so that the warmer and denser gas can be
probed. In addition, they allow deep searches for lines of less
abundant molecules which are important for constraining the
chemistry. Figure~6 shows the spectra obtained toward the source
positions. The velocity-integrated maps are presented in
Fig.~7. Table~7 lists the integrated line intensities in all observed
transitions, together with estimates of the opacity averaged over the
line profile. These are derived from measurements of the same
transition in different isotopes, assuming isotope ratios of ${\rm
[^{13}C] : [^{12}C]=1:65}$ and ${\rm [^{18}O] : [^{17}O] :
[^{16}O]=5:1:2695}$ (Wilson \& Rood \markcite{wil94}1994). Opacities
at line center are much larger, as evidenced by deep self-absorption
features evident in many lines.

The observed line profiles of the optically thick $^{12}$CO, HCO$^+$,
and HCN lines are characterized by $\sim 20$--$30$ km~s$^{-1}$ wide
line wings and deep, narrow ($\sim 1$--2 km~s$^{-1}$) self-absorption
features, in addition to 2--4 km~s$^{-1}$ FWHM line cores. Toward
SMM~3 and 4 the blue asymmetry characteristic of infall is present in
the $^{12}$CO and HCO$^+$ lines (cf.\ Gregersen et~al.\
\markcite{gre97}1997). The same lines toward SMM~1 show symmetric
profiles, however. In optically thin tracers like H$^{13}$CO$^+$,
C$^{18}$O, C$^{17}$O, and H$^{13}$CN only a simple Gaussian line of
$\sim 2$ km~s$^{-1}$ FWHM is seen. From the C$^{17}$O and C$^{18}$O
lines systemic velocities of the sources are derived to be $+8.5$
km~s$^{-1}$ (SMM~1), $+7.6$ km~s$^{-1}$ (SMM~2), $+7.9$ km~s$^{-1}$
(SMM~3), and $+7.9$ km~s$^{-1}$ (SMM~4). The $^{13}$CO lines show
moderate self-absorption and $\sim 10$ km~s$^{-1}$ wide
wings. Many of the other observed transitions show simple Gaussian
line profiles, although some lines (e.g., H$_2$CO $3_{03}$--$2_{02}$
and $3_{22}$--$2_{21}$) are unresolved at the velocity resolution
obtained. Hurt et~al.\ \markcite{hur96b}(1996b) present
observations of these lines at higher spectral resolution, which
reveal moderate self-absorption features. The integrated line
intensities are unaffected, as long as the overall line width is still
comparable to the instrumental resolution.

The $^{12}$CO 4--3 and 6--5 line profiles of Fig.~6 reveal
an interesting hint about the temperature and velocity structure of
the outflowing gas. While the line wings in $^{12}$CO 4--3 decrease
smoothly to more red and blue-shifted velocities, the $^{12}$CO 6--5
wings show secondary maxima at $\sim 7$--9 km~s$^{-1}$ from the
systemic velocity. This indicates that the excitation temperature of
the gas increases with velocity in the outflow, to $\gtrsim 200$ K at
the velocity of the secondary maxima in $^{12}$CO 6--5. In addition, a
larger fraction of the 6--5 line cores is self-absorbed as compared to
the 4--3 lines. The $^{12}$CO 6--5 line profiles toward the low-mass
YSO TMC~1A in Taurus show a similar secondary maximum in the blue line
wing (Hogerheijde et~al.\ \markcite{mrh98a}1998).

The maps of integrated intensity show well-defined cores around
SMM~1, 3, and 4. Again, SMM~2 appears associated with a
condensation in the overall cloud rather than a YSO. Still, SMM~2 does
show $^{12}$CO 6--5 emission with broad line wings, which requires
kinetic temperatures of 80 K or more to be excited, as well as the
secondary maxima noted above. White et~al.\ \markcite{whi95}(1995)
conclude from $^{12}$CO 2--1 mapping that outflow emission permeates
the whole Serpens cloud core region. Probably, the $^{12}$CO 6--5
lines are tracing this same material at the position of SMM~2 with the
same distribution of excitation temperature with velocity. The
$^{12}$CO 4--3, and HCO$^+$ 3--2 and 4--3 maps toward SMM~1, 3, and 4
are resolved with diameters of $20''$--$30''$ (8000--12000 AU), while
H$^{13}$CO$^+$ 3--2 (SMM~1) and HCN 4--3 (SMM~1 and 4) appear
unresolved at the respective beam sizes of $19''$ (7600 AU) and $14''$
(5600 AU). The maximum of the emission in $^{12}$CO 4--3 toward SMM~4
is offset by $\sim 15''$ to the north. The five-point map observed in
$^{12}$CO 6--5 shows a similar offset. These lines probably trace
$T_{\rm kin} \approx 100$ K gas associated with the
outflow. Absorption of a significant fraction of the line profiles by
cold and dense envelope material within $\sim 15''$ from the YSO could
explain the apparent offsets of the peak in the integrated intensity
maps.

\subsection{A model for the molecular line emission}

Section 4.2 described a model for the envelope based on the dust
continuum emission, with $\rho \propto r^{-2}$ and $T_{\rm d} \propto
r^{-0.4}$. Although the inner few hundred AU of the envelope is
relatively warm, 80\% of the material is below the $\sim 30$ K
required to excite many of the observed molecular lines. Therefore, an
important question is if the envelope model can reproduce the observed
line emission. Related issues are the limits set by the data on the
molecular abundances, and on their possible depletion by freezing out
onto dust grains.

We use a Monte-Carlo code recently developed by Hogerheijde \& van der
Tak \markcite{mrh98z}(1998) to solve the non-LTE excitation and line
transfer in a spherically symmetric envelope with the density and
temperature distribution of \S 4.2. An inner radius of 100~AU is
adopted, which does not influence the results.  The core is assumed to
be static with a turbulent width that is constant with radius. FWHM
values of 1.4 km~s$^{-1}$ for SMM~1, 2.1 km~s$^{-1}$ for SMM~3, and
2.0 km~s$^{-1}$ for SMM~4 reproduce the observed FWHM line width of
the C$^{17}$O, C$^{18}$O, and H$^{13}$CO$^+$ lines. The effect of
neglecting any systematic velocity field, like infall, is to increase
the optical depth for material close to the center, thereby decreasing
the integrated intensity of optically thick lines of $^{12}$CO,
HCO$^+$, and HCN (cf.\ Hogerheijde \markcite{phd98}1998; Hogerheijde
\& van der Tak \markcite{mrh98z}1998). Because of the imperfect
thermal coupling between gas and dust, and the cooling of the gas
through line radiation, the gas temperature may be lower than that of
the dust. Self-consistent models by Ceccarelli et~al.\
\markcite{cec96}(1996) and Doty \& Neufeld \markcite{dot97}(1997)
suggest $T_{\rm kin}= (0.6$--$0.8) \times T_{\rm dust}$.

The molecular abundances are a free parameter of the model, in
addition to the exact relation between $T_{\rm kin}$ and $T_{\rm
dust}$. Standard isotopic ratios are assumed (Wilson \& Rood
\markcite{wil94}1994), as well as an ortho-to-para ratio of 3:1 for
H$_2$CO and C$_3$H$_2$. The abundance of $^{12}$CO is fixed at the
`standard' value of $10^{-4}$ with respect to H$_2$, except for the
possibility of depletion onto dust grains. In our model, CO is allowed
to freeze out on dust grains when the temperature drops below the
sublimation temperature of 20 K (Sandford \& Allamandola
\markcite{san90}1990, \markcite{san93}1993). Absorption studies toward
nearby infrared sources provide evidence for solid CO in the Serpens
cloud (Chiar et~al.\ \markcite{chi94}1994), with $\sim 40\%$ of the
total CO column frozen out onto grains. The optically thin C$^{18}$O
2--1 and C$^{17}$O 3--2 lines set limits both on the depletion and the
relation between $T_{\rm kin}$ and $T_{\rm dust}$. Assuming $T_{\rm
kin}=T_{\rm dust}$ and no CO depletion, the model overestimates the
C$^{18}$O 2--1 and C$^{17}$O 3--2 intensities by factors of 1.5--2.0
for the different sources. A better fit is found by adopting $T_{\rm
kin} < T_{\rm dust}$ or by depleting CO in the coldest region of the
cloud. A lower limit of $T_{\rm kin}=(0.7$--$0.8) \times T_{\rm dust}$
is found if CO is undepleted. Alternatively, the observed intensities
are reproduced if CO is depleted by a factor of 3--10 for SMM~1, 2--4
for SMM~3, and 2--4 for SMM~4 in regions with $T_{\rm kin}<20$ K,
using $T_{\rm kin}= T_{\rm dust}$. This provides a slightly better fit
to the data, since C$^{18}$O 2--1, which traces 16 K gas, is more
strongly affected than C$^{17}$O 3--2, which traces 30 K gas. These
depletion factors are maximum values; less depletion is found for
$T_{\rm kin} <T_{\rm dust}$. These results depend on the adopted dust
emissivity, but the depletion would be much larger only if the
emissivity were overestimated by a large factor.  The data therefore
show that molecules are not strongly depleted in these sources (cf. \S
6.2).

Table 8 lists the molecular abundances derived from the observations
using the envelope model. For simplicity, the calculations assume
$T_{\rm kin} = T_{\rm dust}$ and the derived depletion factors in
regions with $T_{\rm kin}<20$ K. None of these assumptions influences
the results by more than a factor of 2, especially since most observed
transitions only trace material with $T_{\rm kin}>30$ K. The inferred
abundances (`envelope' column of Table 8) are compared to results for
another class~0 YSO, IRAS 16293$-$2422 (van Dishoeck et~al.\
\markcite{evd95}1995). The values of most species agree within a
factor of 2 to 3. SiO and SO have smaller inferred values toward the
Serpens sources, because our single-dish beams did not contain the
emission peak apparent in the interferometer images (cf.\ 5.1). Other
differences are found toward SMM~1, where the abundances of CN,
HC$_3$N, and C$_3$H$_2$ are larger by an order of magnitude compared
to IRAS 16293$-$2422. The inferred abundance of C~I is highly
uncertain. With its critical density of only $10^3$ cm$^{-3}$, the
[C~I] $^3$P$_1$--$^3$P$_0$ line is likely to trace the low-density
surface of the entire Serpens cloud, where the interstellar radiation
field has returned much of the carbon in the atomic phase. White
et~al.\ \markcite{whi95}(1995) reach that same conclusion from the
lack of [C~I] emission maxima at the positions of the submillimeter
cores.

Most lines are well reproduced for these envelope parameters. Notable
exceptions are the optically thick $^{12}$CO lines, for which the
treatment of the velocity field is too simple and where the outflow
also contributes significantly, and high excitation lines like
$^{13}$CO 6--5 and H$_2$CO $3_{22}$--$2_{21}$. This indicates the
presence of more warm, $T_{\rm kin} \sim 100$ K, material than can be
accounted for by the model. The line ratio of H$_2$CO
$3_{03}$--$2_{02}$ over $3_{22}$--$2_{21}$ is a very sensitive
diagnostic of kinetic temperature (Mangum \& Wootten
\markcite{man93}1993; Jansen, van Dishoeck, \& Black
\markcite{djj94}1994; Hurt et~al.\ \markcite{hur96b}1996b). Our data
imply a gas density in excess of $10^6$ cm$^{-3}$, and a kinetic
temperature of 50--200 K, consistent with the findings of Hurt et~al.\
\markcite{hur96b}(1996b). Adopting the parameters derived by these
authors, the excess $^{13}$CO 6--5 emission is reproduced by column
densities of $7\times 10^{21}$ cm$^{-2}$ toward SMM~1, $1\times
10^{21}$ cm$^{-2}$ toward SMM~3, and $5\times 10^{20}$ cm$^{-2}$
toward SMM~4. This material, less than 1\% of the total envelope mass,
may be associated with the inner few hundred AU of the envelopes where
the temperature exceeds the adopted power-law distribution, or with
the outflows, as suggested by the $^{13}$CO and $^{12}$CO line
wings. It contributes no more than 10\% to the total continuum flux,
and, if confined to the inner 1000 AU, could explain the point source
fluxes required to fit the interferometer continuum observations of \S
4.2. Could this material dominate the emission in the other observed
molecular lines? Table 8 (`warm gas' column) lists the molecular
abundances derived under the assumption that this component is
responsible for all observed emission except CO. All values are larger
by an order of magnitude compared to those found from the `cold
envelope', indicating that the warm gas may contribute significantly
to the observed intensities, but only if the abundances are much
enhanced.

On the 300--4000 AU scales sampled by the interferometer,
the envelope model predicts line intensities within a factor of 5 of
the observed values for most lines. To derive these intensities, the
model data have been sampled at the same $(u,v)$ positions as the
observations.  In \S\S 5.1.1--5.1.4 masses were derived from the
$^{13}$CO and C$^{18}$O lines observed with OVRO which were 30 times
smaller than the values derived in \S 4.2 from the dust emission. The
results of Table~8 show that such low fractions of recovered emission
can be mostly explained by the envelope parameters. This good
agreement indicates that there are no strong abundance changes on
small scales. However, a more realistic treatment of the velocity
field and possible variations from spherical symmetry on small scales
is required before stronger abundance constraints can be
derived.


\section{Discussion}

\subsection{Envelopes and disks around class~0 and I YSOs}

This section compares the density structure of the Serpens class~0
YSOs studied in this paper to results for other class~0 objects and
more evolved class~I sources. The inferred density structure for the
envelopes around the Serpens YSOs as a radial power law with a slope
of $-2.0\pm 0.5$ agrees well with previous results for class~0 YSOs
and theoretical expectations. Zhou et~al.\ \markcite{zho93}(1993) and
Choi et~al.\ \markcite{cho95}(1995) find that molecular line profiles
observed toward the class~0 YSOs B335 and IRAS~16293$-$2422 can be
accurately reproduced using the inside-out collapse model of Shu
\markcite{shu77}(1997). This model predicts density power-law indices
between $-1$ and $-2$. Recent millimeter-continuum mapping
observations of dense cores with and without embedded stars have shown
that the former have density distribution with $\rho \propto r^{-p}$
with $p\approx 2$, while the latter have a significantly flatter
distribution in the inner few thousand AU (Ward-Thompson et~al.\
\markcite{war94}1994; Andr\'e, Ward-Thompson, \& Motte
\markcite{and96}1996; and Motte, Andr\'e, \& Neri
\markcite{mot98}1998). Our interferometric millimeter-continuum
observations directly trace these size scales, and confirm that the
envelopes around class~0 YSOs are strongly centrally concentrated.

In two recent papers (Hogerheijde et~al.\ \markcite{mrh97a}1997,
\markcite{mrh98a}1998), we investigated a sample of nine class~I YSOs in
Taurus through a data set which is very similar to that presented
here.  The millimeter continuum emission toward at least two-thirds of
these class~I sources is dominated by an unresolved component,
presumably a disk, carrying 30\%--75\% of the total 1~mm flux. The
point-source flux required in \S 4.2 to fit the interferometer
continuum observations of the SMM~1, 3, and 4 amounts to only
20\%--30\% of the total 1.1~mm flux. This indicates that any
circumstellar disk makes a much smaller relative contribution to the
total flux of class~0 sources than for class~I objects, especially
since a sizeable fraction of this point-source emission may also be
attributable to the warm, inner 100 AU of the envelope.  No limits on
disk masses can therefore be obtained for the Serpens class~0 sources
from the present data. 

Significant, resolved emission around the Taurus class~I sources was
only detected toward L1551~IRS~5 and L1527~IRS. The visibility fluxes
of the latter source, which is sometimes referred to as a class~0
source, closely resemble those of the Serpens class~0 YSOs, but its
total envelope mass is only 0.03 $M_\odot$, and the signal-to-noise is
insufficient to fit an envelope model as in \S 4.2. The envelopes
around all Taurus class~I sources, including L1527~IRS, contain less
than 50\% of the estimated stellar mass, consistent with their higher
age. Because of their lower masses, these envelopes of class~I sources
are much better traced by molecular lines than by continuum emission.
HCO$^+$ 1--0, 3--2, and 4--3 single-dish observations of the Taurus
class~I sources showed that their envelopes are well described by a
power-law density structure, similar to that derived here for the
Serpens class~0 objects. From the integrated intensities observed by
Hogerheijde et~al.\ \markcite{mrh97a}(1997), the power-law slope is
constrained to lie between $-1$ and $-3$. These values are consistent
with predicted values of $-1$ to $-2$ for a collapsing cloud (Shu
\markcite{shu77}1977), and with the values for the class~0 sources
found here. By including the complete observed line profile in the
model fit, much more accurate constraints on age and envelope density
structure can be obtained, which will allow tests of other collapse
models (cf.\ Zhou \markcite{zho92}1992, \markcite{zho95}1995; Zhou
et~al.\ \markcite{zho93}1993; Walker, Narayanan, \& Boss
\markcite{wal94}1994; Choi et~al.\ \markcite{cho95}1995; Hogerheijde
\markcite{phd98}1998; Hogerheijde \& van der Tak
\markcite{mrh98z}1998).

The Serpens YSOs studied in this paper have formed within a projected
distance of 25,000 AU of each other. The detection of widespread
emission from moderate $J$ lines of CO with broad profiles (White
et~al.\ \markcite{whi95}1995) indicates that material throughout the
Serpens cloud is influenced by the outflows. It is therefore
remarkable that our data, which probe the 16,000 AU diameter envelopes
on 1000 AU scales, are consistent with the density structure predicted
for the formation of an isolated star (Shu \markcite{shu77}1977; \S
4.2). This suggests that the influence of the outflows on the cloud
structure is largely confined to lower density material, leaving the
dense cores mostly unaffected. Alternatively, their envelopes may
reflect the original, relatively isolated state if all three YSOs have
formed within an interval shorter than the dynamic time scale of the
outflows of $\sim 8000$ yr as estimated from the $^{12}$CO 2--1 data
of White et~al.\ \markcite{whi95}(1995). However, the projected
location of SMM~3 at the edge of the outflow driven by SMM~4, and the
small age suggested by the highly collimated nature of SMM~3's outflow as
traced by HCN 1--0, may indicate induced star formation. Single-dish
mapping of the entire cloud on $10''$--$15''$ scales, which are well matched
to the interferometer observations, are required to further investigate
the influence of star formation on the cloud structure. Such
observations may also shed further light on the nature of the
apparently starless core SMM~2 and the $\sim 30$ low-mass cores
identified recently by Testi \& Sargent \markcite{tes98}(1998) from an
extensive interferometric survey of the Serpens cloud core.

\subsection{Molecular abundances in class~0 envelopes}

The model results of \S 5.3 indicate that CO is depleted by factors of
2--10 in the cold regions of the envelopes where $T_{\rm kin}<20$ K,
its sublimation temperature, but not in the warmer inner regions of
the envelope. Such a level of depletion is small compared to the value
of 10--20 inferred for the bulk of the envelope of two other class~0
YSOs, NGC 1333 IRAS 4A and B (Blake et~al.\
\markcite{gab95}1995). Submillimeter observations of the latter
sources indicate envelope masses of 9 and 4 $M_\odot$ in a $20''$
beam, adopting a distance of 350 pc, comparable to or larger than
those of the Serpens sources. Based on a detailed excitation analysis,
Blake et~al.\ conclude that the abundances of all molecules including
CO are depleted by a factor of 10--20 when the dust emission is used
as a reference to determine the total column density. Blake et~al.\
used a value for the dust emissivity which was smaller by a factor of
2 than the value adopted by us. If the same value had been assumed,
their derived depletion would be even larger, increasing the
difference with the depletion we find toward the Serpens sources. This
highlights the difficulty in deriving and interpreting molecular
depletion (cf.\ Mundy \& McMullin \markcite{lgm97}1997). The large
difference in molecular depletion between class~0 YSOs which are
otherwise very similar suggests that the quiescent evolutionary phase
characterized by heavily depleted abundances may be relatively
short-lived. In addition, the local star-formation density may be
important. The largest levels of CO depletion are inferred for the
relatively isolated NGC~1333 IRAS~4 object; lower values are found
toward SMM~1, which is located in the northwest of the Serpens region,
while the lowest values are inferred for SMM~3 and 4, at the very
center of the densely star forming cloud. Possibly the high density of
star formation activity inhibits CO depletion, suggesting that the
chemistry may be more sensitive to the environment than the physical
structure of the envelopes (cf.\ previous section).

The line profile shapes toward NGC~1333 IRAS 4A and B resemble those
of the Serpens sources. Blake et~al.\ conclude that the HCN 4--3
profile, which is significantly broader than, e.g., HCO$^+$ 4--3,
indicates that HCN may be enhanced in the outflow. Our HCN 4--3
spectra show this same trend.  In the interferometer beam, H$^{13}$CN
1--0 peaks toward the northern outflow lobe of SMM~3, supporting the
interpretation of larger abundances in the outflow. Enhanced
abundances of volatile species like H$_2$CO and CH$_3$OH toward
NGC~1333 IRAS 4A and B are attributed by Blake et~al.\ to grain-grain
collisions in the velocity-shear zones surrounding the outflow. These
collisions heat the dust grains transiently and evaporate the ice
mantles. Thus, the dynamic interaction of the outflow with the
surrounding envelope could explain the bright emission of some species
outlining the walls of the outflow cavities of YSOs.


\section{Summary}

Continuum and molecular-line emission from four protostellar
candidates in the Serpens Molecular Cloud (SMM~1 = FIRS~1, SMM~2,
SMM~3, and SMM~4) has been used the provide the first comprehensive
picture of the physical and chemical structure on 1000--10,000 AU
scales in a clustered region.  We confirm the nature of SMM~1, 3, and
4 as deeply embedded, class~0 YSOs with more than 50\% of their total
mass still in a circumstellar envelope. The density in these envelopes
is well described by a power-law with slope $-2.0\pm 0.5$. This agrees
with theoretical predictions for an isolated cloud core, even though
the SMM~1, 3, and 4 are in close proximity of each other, and
outflowing gas permeates the whole Serpens core. The characteristics
of, at least, the early phases of star formation therefore do not seem
to be very different between isolated and more clustered
environments. Large degrees of molecular depletion by freezing-out of
molecules onto grains have been inferred for some class~0 YSOs, but
the low levels of depletion found toward the Serpens sources suggests
this phase may be short-lived and/or influenced by the surrounding
star-forming activity. SMM~2 is most likely a warm cloud condensation
without a central object since it shows no compact continuum or
molecular line emission.  The main conclusions of this paper can be
summarized as follows:

1.~The interferometric observations of SMM~1, 3, and 4 in 3.4--1.4 mm
  dust continuum emission are excellent tools to derive the structure
  of the envelopes surrounding these class~0 sources. They are
  characterized by intense, resolved ($>2''=800$ AU) emission which
  can be described by a density distribution that follows a radial
  power law with index $-2.0\pm 0.5$. Such a density distribution
  agrees well with model predictions for the earliest phases of
  protostellar collapse (e.g., Shu \markcite{shu77}1977). The
  temperature distribution is based on self-consistent model results,
  and fits to the millimeter and far-infrared SED. At a representative
  radius of 1000 AU, dust temperatures of 27 K (SMM~1), 24 K (SMM~3),
  and 20 K (SMM~4) are found. The mass contained within the envelopes
  is constrained to 8.7 $M_\odot$ for SMM~1, 3.0 $M_\odot$ for SMM~3,
  and 5.3 $M_\odot$ for SMM~4. To fit the flux on long baselines, a
  point source with 20\%--30\% of the total flux is required,
  corresponding to emission originating from radii $<100$
  AU. Observations at higher spatial resolution are required to
  investigate how much of this flux is due to a circumstellar disk and
  how much to the dense inner 100~AU of the power-law envelope. Such
  observations for other class~0 YSOs suggests the envelope dominates
  the emission even on small scales.

2.~The aperture-synthesis molecular line observations reveal
  $5''$--$10''$ (2000--4000 AU) diameter cores around SMM~1, 3, and 4
  in C$^{18}$O, $^{13}$CO, H$^{13}$CO$^+$, and H$^{13}$CN. The HCO$^+$
  and HCN emission traces the walls of the outflow cavities, while SiO
  and SO emission originates from material shocked by the
  outflows. Molecular species and transitions are thus identified which
  selectively probe different physical regions of the envelope,
  offering diagnostic tools for future studies of the protostellar
  environment.

3.~Higher excitation single-dish lines show broad outflow wings and
  deep, narrow self-absorption features, in addition to Gaussian line
  cores of 2--3 km~s$^{-1}$ FWHM. Monte-Carlo calculations of the
  molecular excitation and line transfer show that the envelope model
  derived from the dust emission can successfully reproduce the
  observed single-dish and interferometric line intensities, except
  for the $\lesssim 1\%$ of warm material traced in the $^{12}$CO,
  $^{13}$CO 6--5, and some of the H$_2$CO lines. The CO abundance
  appears to be decreased by a factor of 2--6 in the regions of the
  envelope where the gas temperature is 20 K or less ($r>2000$ AU),
  owing perhaps to freezing-out onto grains. The derived molecular
  abundances in the warmer gas ($T_{\rm kin}>20$ K) are comparable to
  those found toward other class~0 YSOs, but some species may be
  enhanced toward SMM~1. The model predicts interferometric line
  fluxes which are within a factor of a few of the observed values,
  indicating that there are no large abundance variations on 300--4000
  AU scales.  However, a more realistic treatment of the velocity
  field and possible deviations from spherical symmetry have to be
  included before firmer limits can be placed on the abundances.


\acknowledgments

The authors are grateful to the staffs of the JCMT, CSO, and OVRO
telescopes for their assistance. Remo Tilanus and Fred Baas are
thanked for carrying out part of the JCMT observations. Lee Mundy and
Huib Jan van~Langevelde are acknowledged for useful
discussions. Floris van der Tak kindly assisted in deriving the
self-consistent temperature structure of the envelopes. M.~R.~H. is
indebted to the Caltech Divisions of Geological and Planetary Sciences
and Mathematics, Physics and Astronomy, and the Owens Valley Radio
Observatory for their hospitality, and to the Netherlands Organization for
Scientific Research (NWO) and the Leids Kerkhoven-Bosscha Fonds for
travel support. Research in Astrochemistry in Leiden is supported by
NWO/NFRA through grant no.\ 781--76--015. G.~A.~B. gratefully
acknowledges support provided by NASA grants NAG5--4383 and
NAG5--3733. A critical reading by Xander Tielens and many valuable
comments by the referee helped to improve the presentation.






\newpage

\figcaption[fig1.ps]{Cleaned, uniformly weighted images of continuum
emission at $\lambda=3.4$, 3.2, 2.7, and 1.4 mm of SMM~1, 2, 3, and 4
observed with OVRO. Contour levels are drawn at 3$\sigma$ intervals
(at 3.4, 3.2, 2.7, 1.4~mm: 6, 9, 18, 48 mJy~beam$^{-1}$ for SMM~1; 4,
$\ldots$, 6, 90 mJy~beam$^{-1}$ for SMM~2; 6, $\ldots$, 12, 150
mJy~beam$^{-1}$ for SMM~3; and 6, 9, 9, 45 mJy~beam$^{-1}$ for
SMM~4). Synthesized beam sizes are indicated in each panel. Note the
much smaller depicted region at 1.4 mm.
\label{f1}}

\figcaption[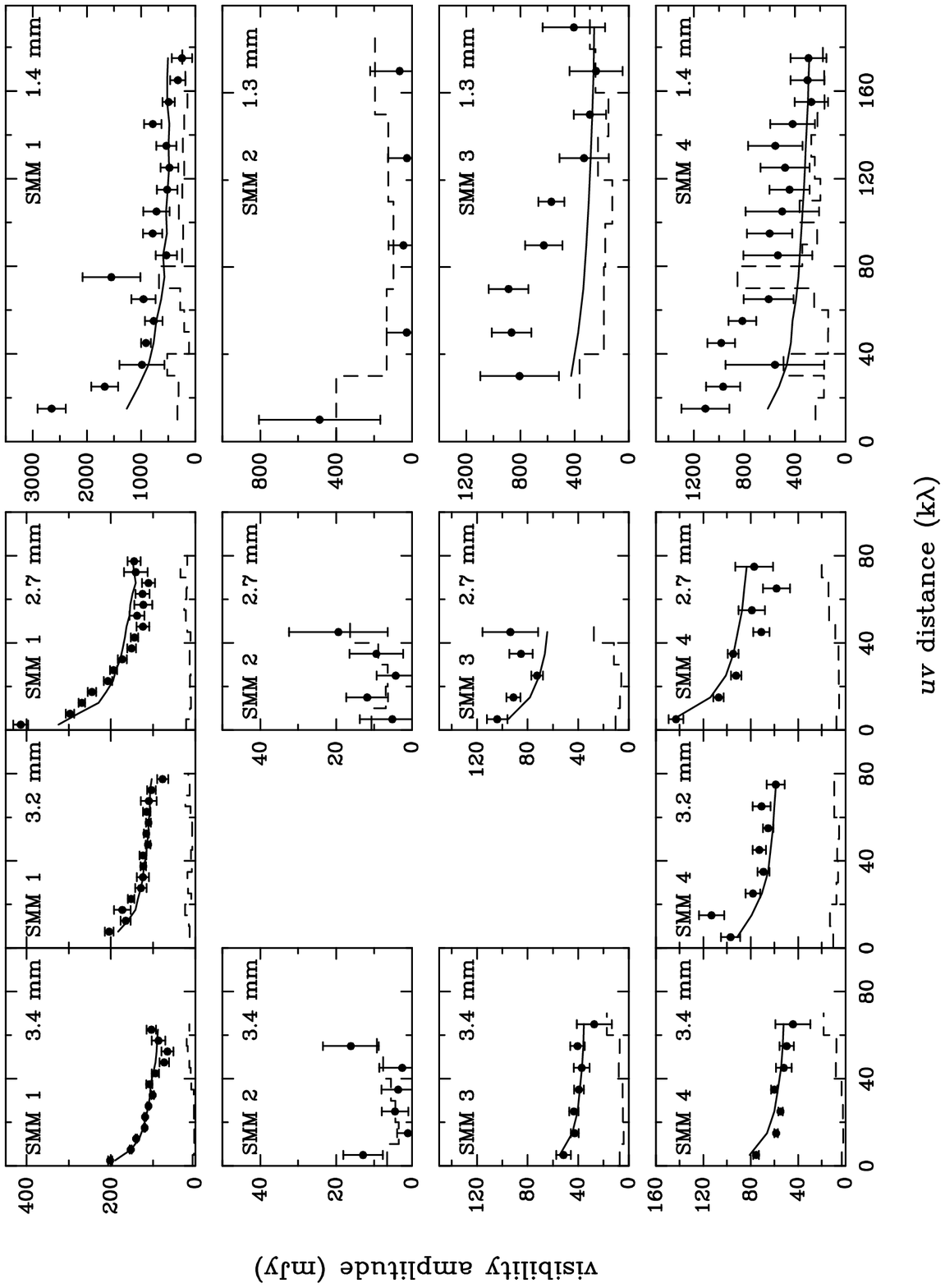]{Vector-averaged visibility amplitudes of observed
continuum emission at 3.4, 3.2, 2.7 and 1.4 mm in mJy as functions of
projected baseline length in k$\lambda$. The data are plotted as
filled symbols, together with 1$\sigma$ error bars. The dashed
histogram shows the zero-signal expectation value. The solid lines are
best-fit models as described in \S 4.2, for a density power-law slope
of $-2.0$ and point source fluxes of \S 4.2. These model curves are
not smooth because of incomplete sampling of the $(u,v)$ plane in the
original data set. 
\label{f2}}

\figcaption[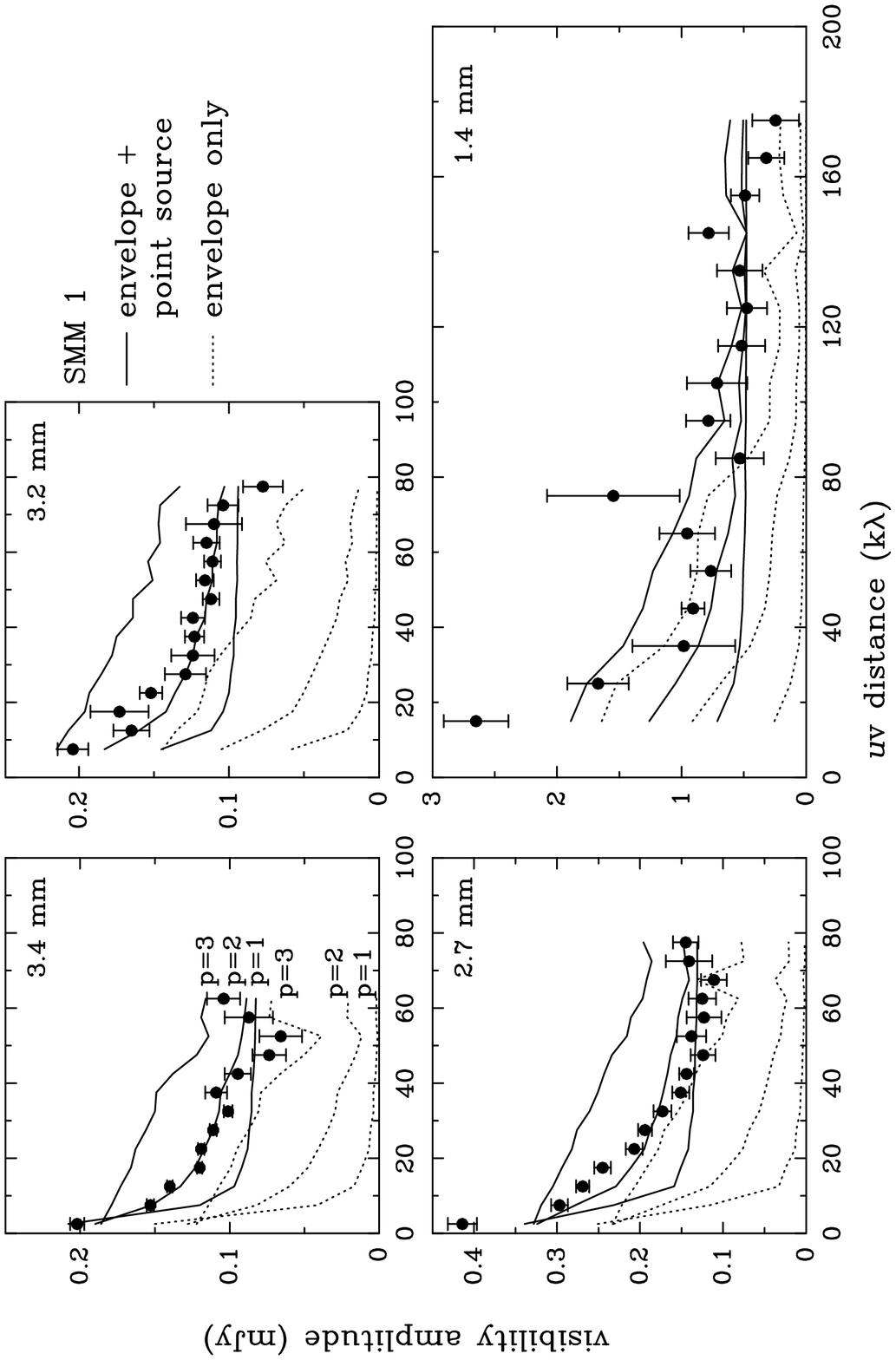]{Comparison of model visibility amplitudes to
observations of SMM~1. The observed vector-averaged amplitudes are
indicated by the filled symbols and their 1$\sigma$ error bars. Dotted
lines are models without a central point source, and density power-law
slopes of $-1.0$ ({\it lower curve in each panel\/}), $-2.0$ ({\it
middle curve\/}), and $-3.0$ ({\it upper curve\/}). Solid lines are
models with a point source flux of 0.13 Jy at 2.7 mm and a spectral
slope of 2.0. 
\label{f3}}

\figcaption[fig4a.ps,fig4b.ps]{({\it a\/}) Cleaned, naturally weighted
images of molecular line emission observed with OVRO toward
SMM~1. Contours are drawn at 3$\sigma$ intervals of 10 ($^{13}$CO), 4
(C$^{18}$O, SiO, HCN, H$^{13}$CN), 1 (HCO$^+$, SO), and 8
(H$^{13}$CO$^+$, N$_2$H$^+$) K~km~s$^{-1}$. The synthesized beam size
is indicated in each panel. The dashed circle shows the primary beam
size. The arrows in the HCN panel indicate the position angle of the 6
cm radio jet of Rodr\'{\i}guez et~al.\ 1989. ({\it b\/}) Same, for
SMM~2, 3, and 4. Contour levels are 1.5 (C$^{18}$O, H$^{13}$CO$^+$),
2.0 ($^{13}$CO, HCO$^+$, HCN), 2.5 (SiO, H$^{13}$CN), and 3.0
(N$_2$H$^+$) K~km~s$^{-1}$.  The images have been cleaned separately,
and mosaicked afterwards. The arrows in the HCN panel indicate the
position angles of the H$_2$ jets (from Herbst et~al.\ 1997 and Eiroa
et~al.\ 1997)\label{f4}}

\figcaption[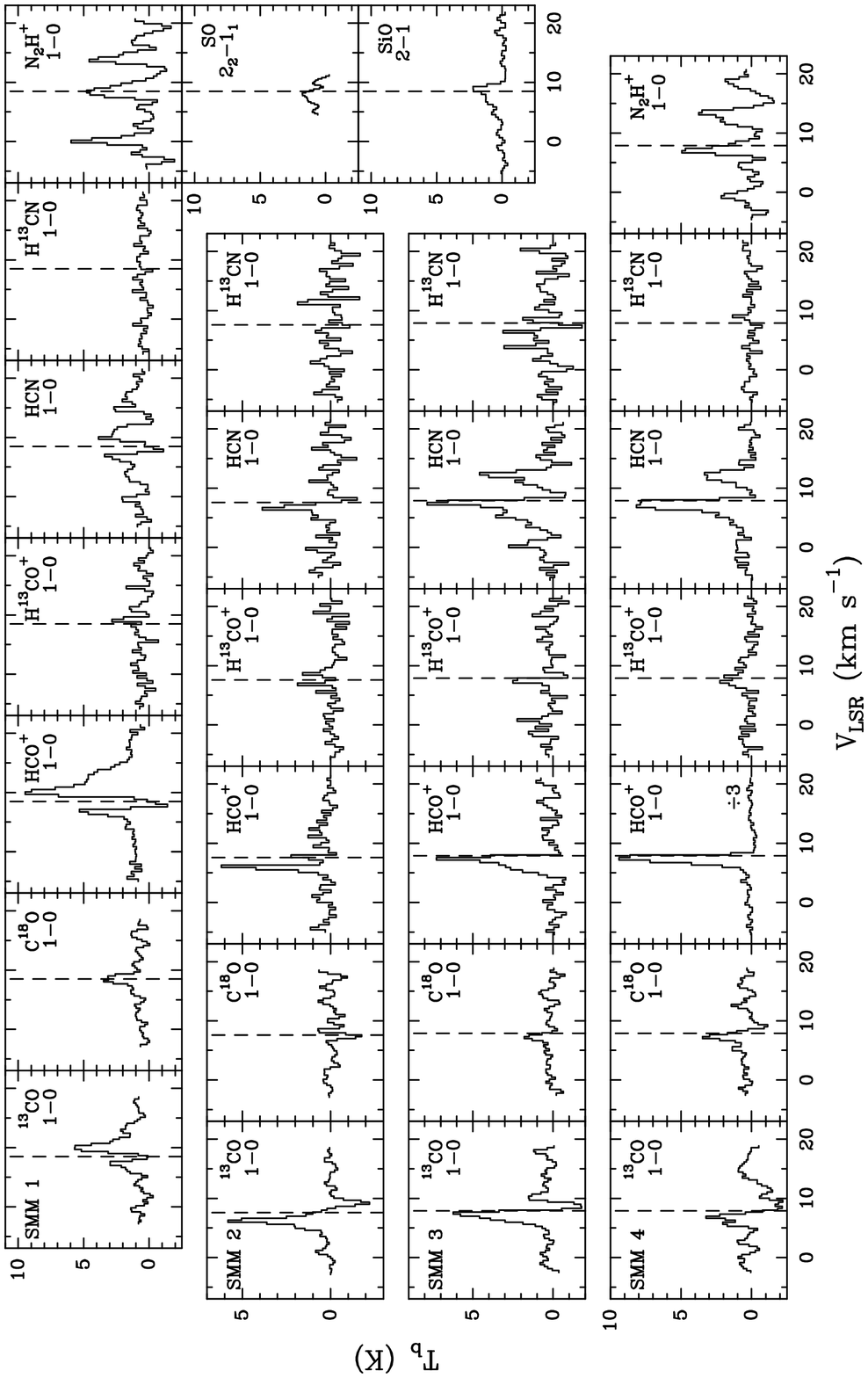]{Spectra obtained within a synthesized beam from
the emission maximum of molecular lines observed with OVRO. Vertical
scale is brightness temperature $T_{\rm b}$, horizontal scale is
velocity $V_{\rm LSR}$. The vertical dashed line indicates the
systemic velocity $V_0$ of the sources.\label{f5}}

\figcaption[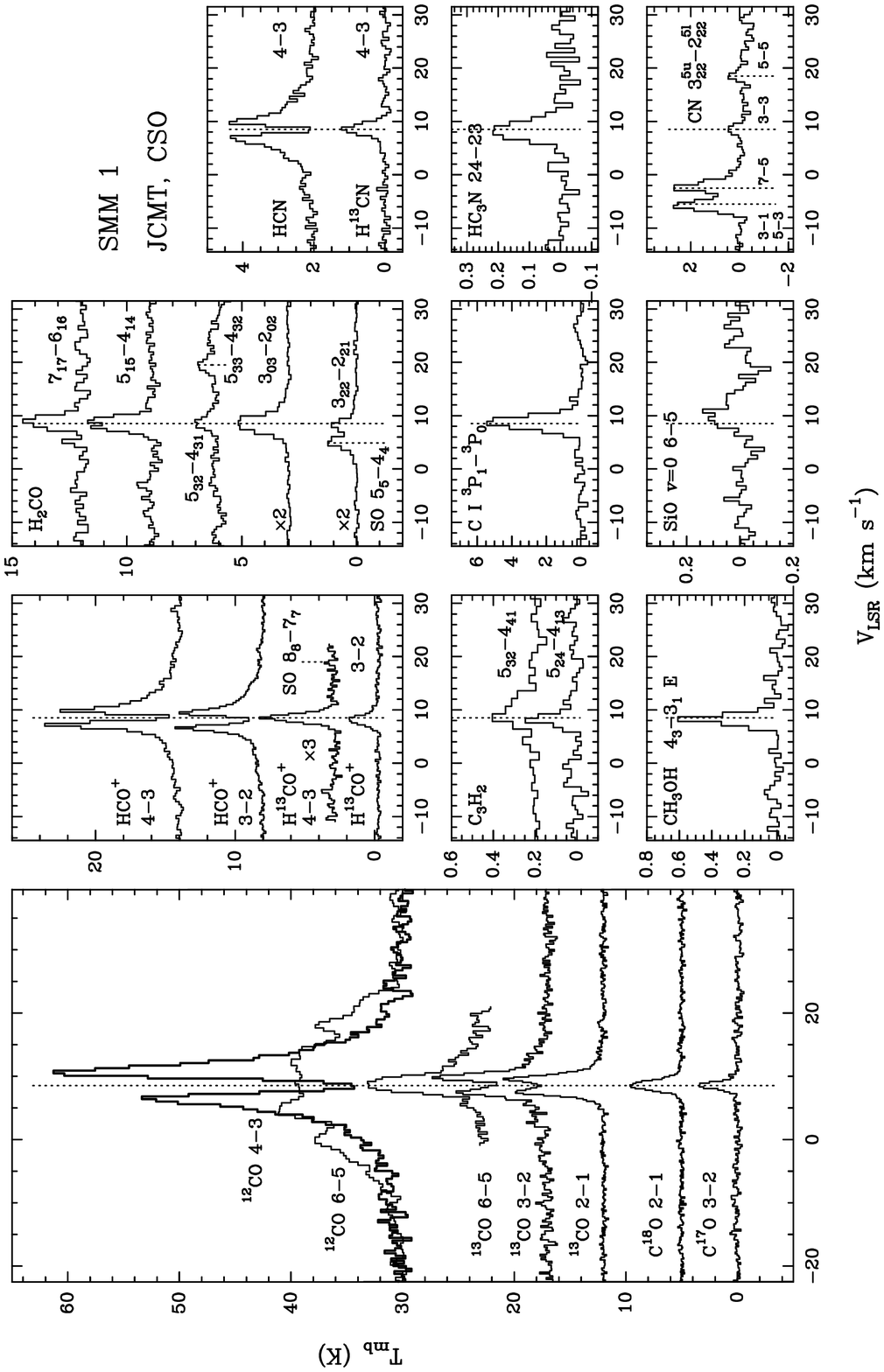,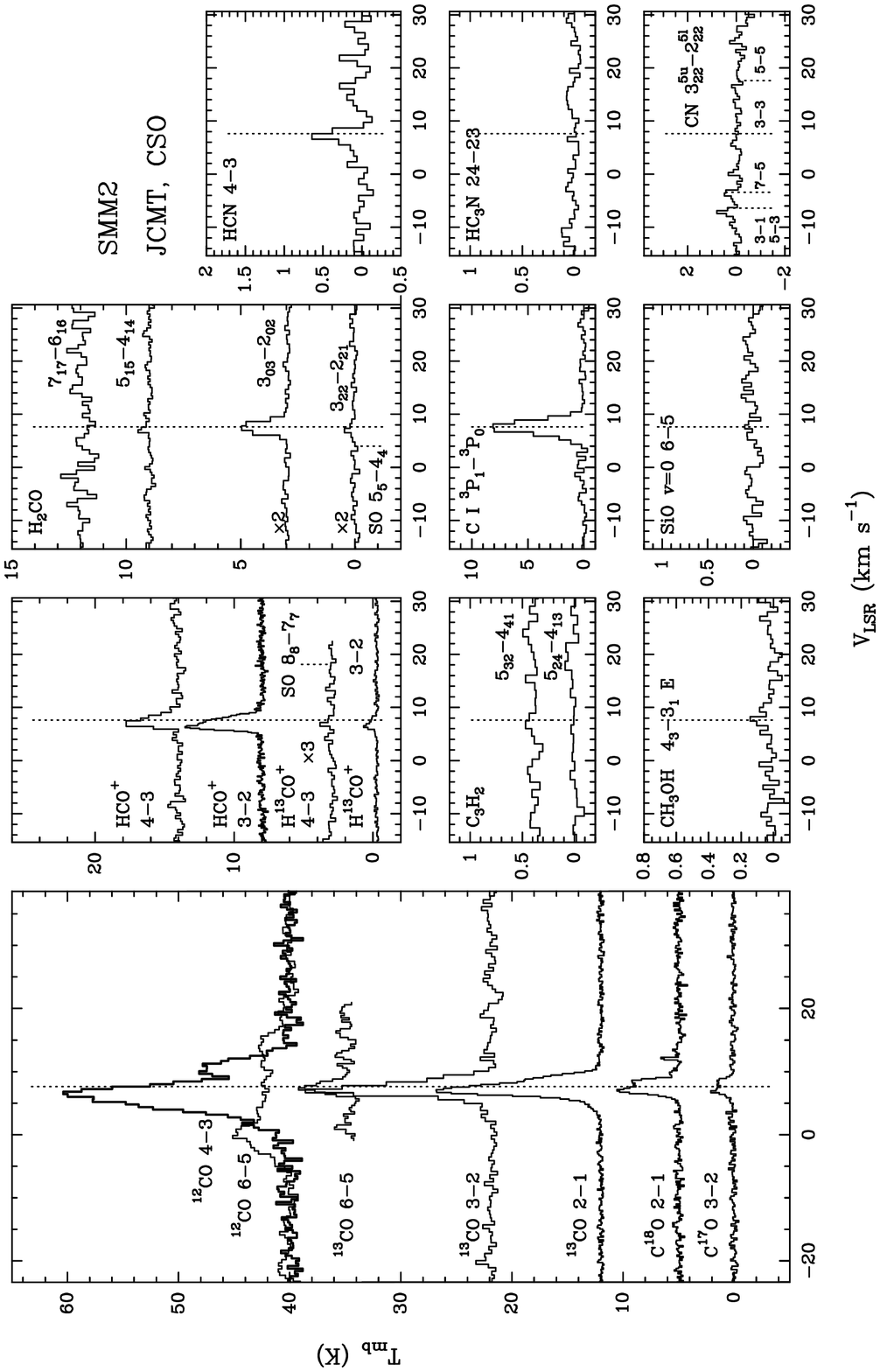,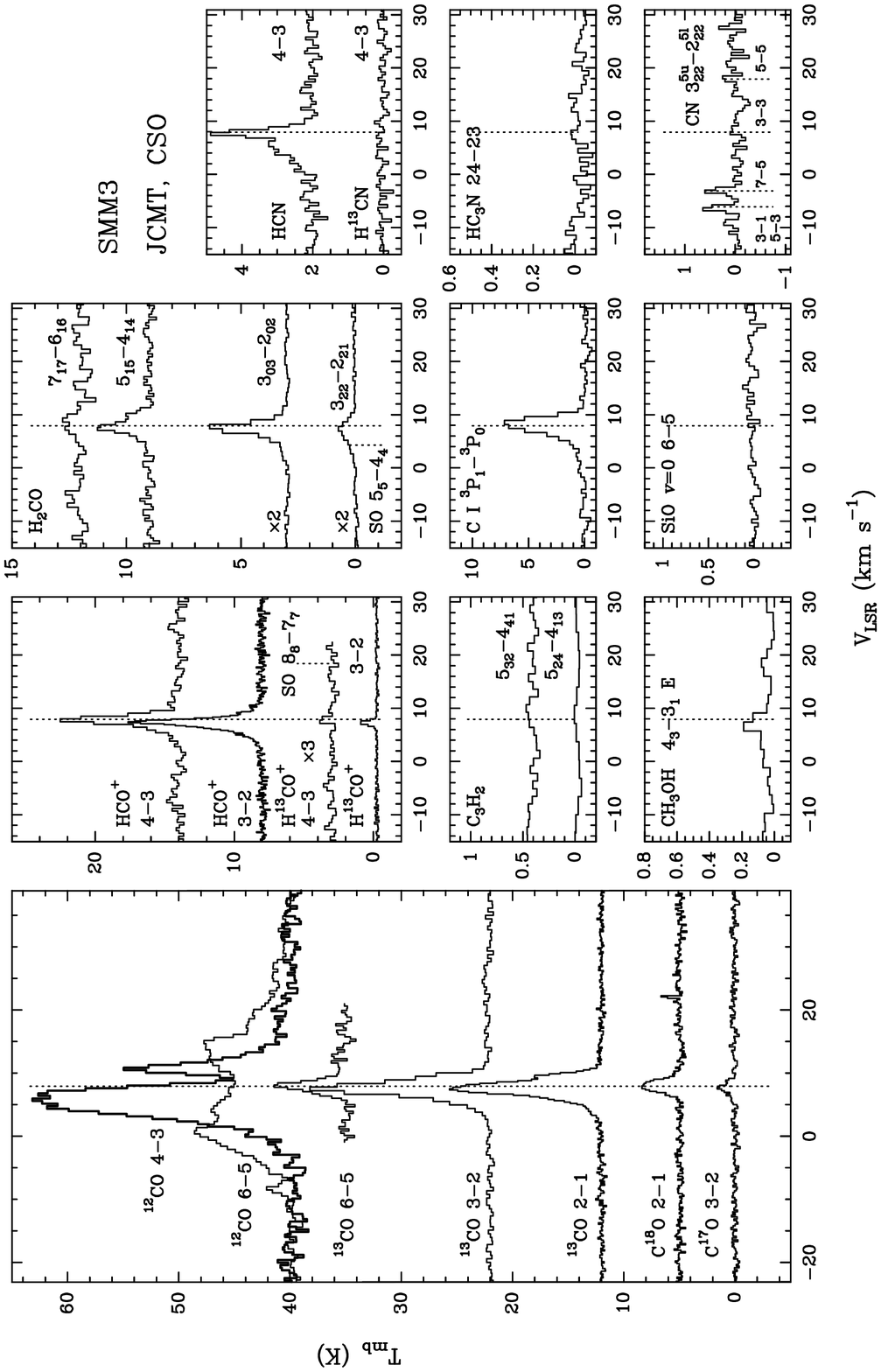,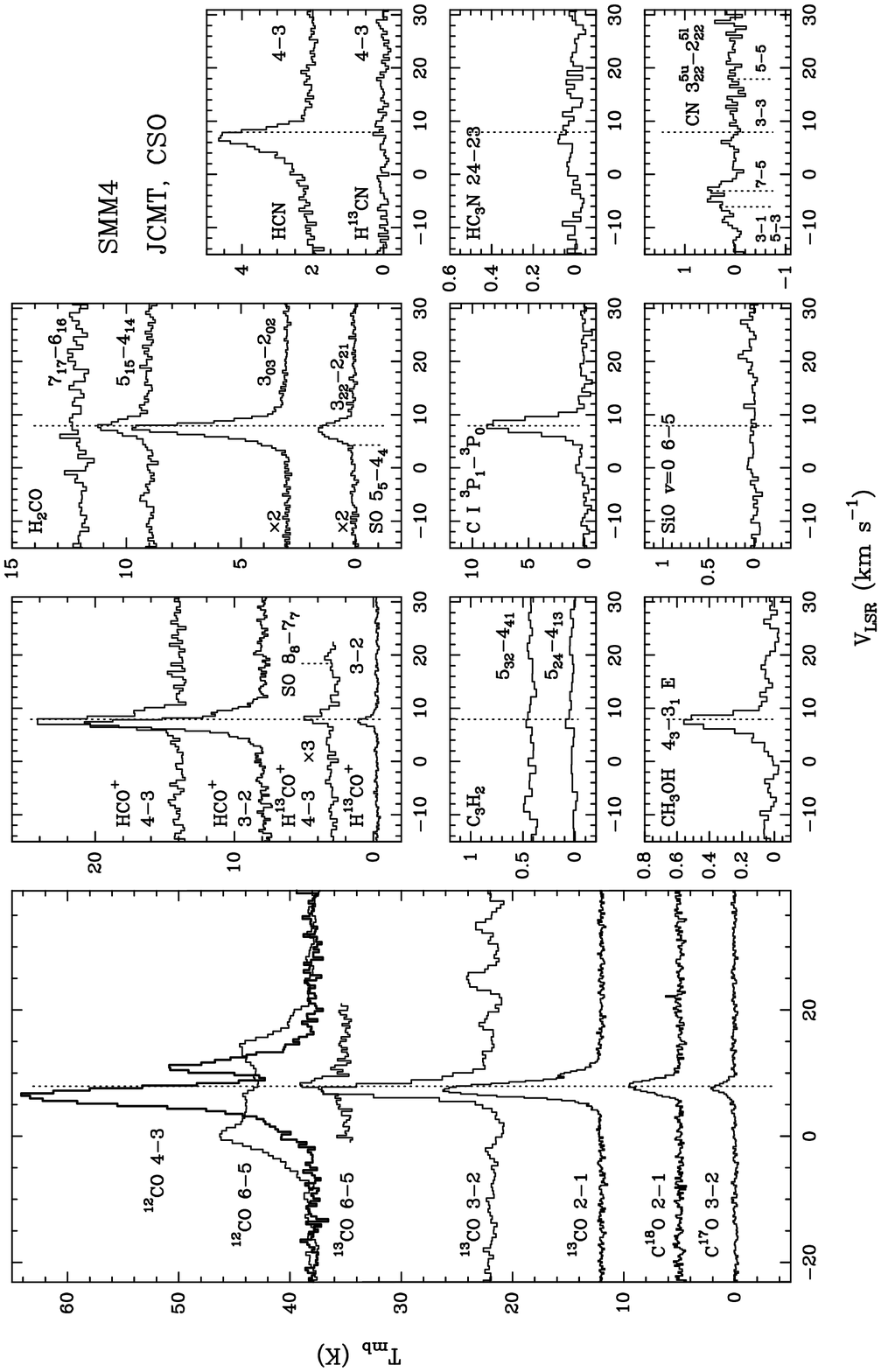]{Spectra obtained in
the single-dish beams. Vertical scale is main-beam temperature $T_{\rm
mb}$, horizontal scale is velocity $V_{\rm LSR}$. The vertical dotted
line indicates the systemic velocity of the source. ({\it a\/}) SMM~1,
({\it b\/}) SMM~2, ({\it c\/}) SMM~3, ({\it d\/}) SMM~4. 
\label{f6}}

\figcaption[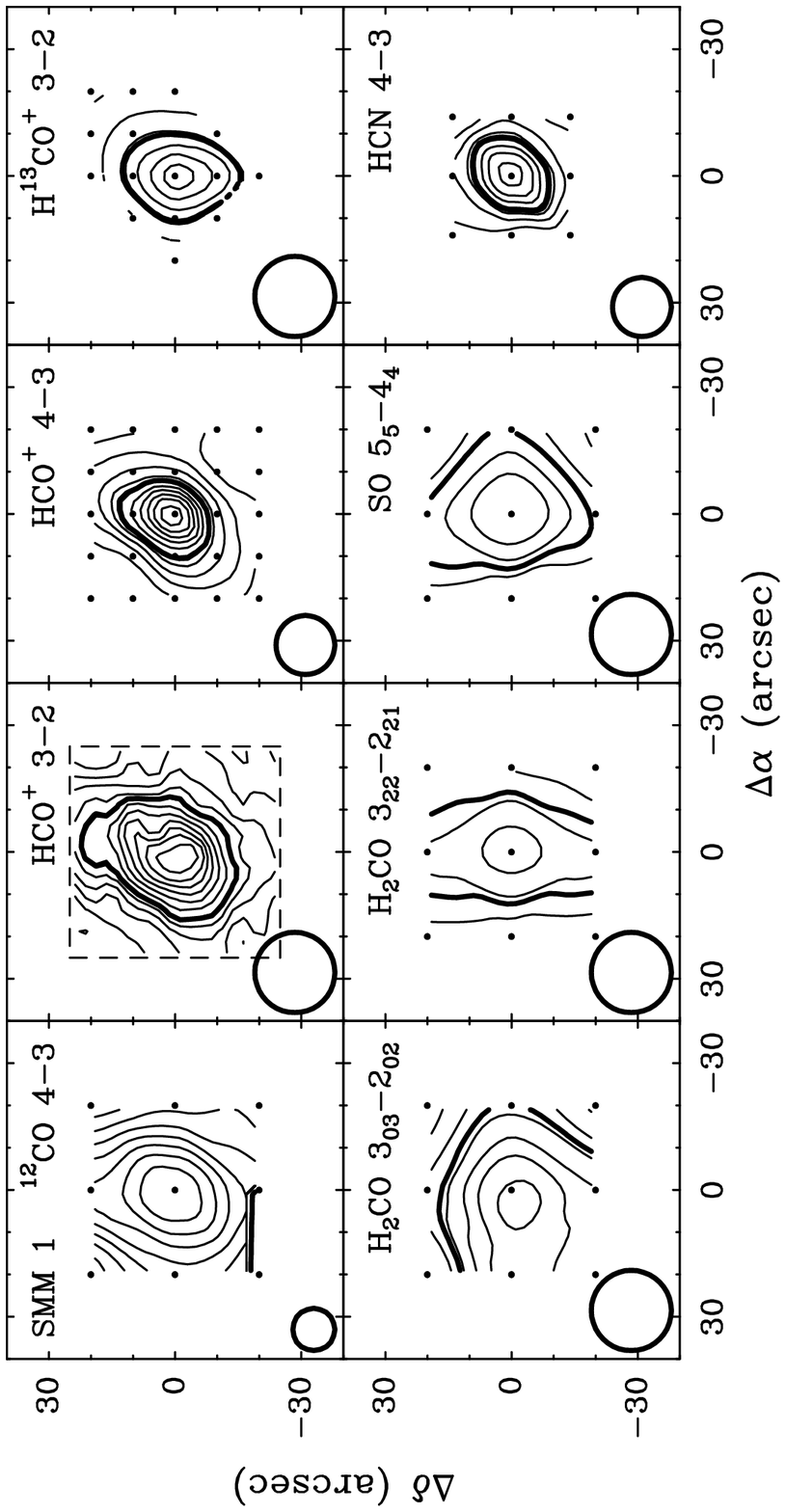,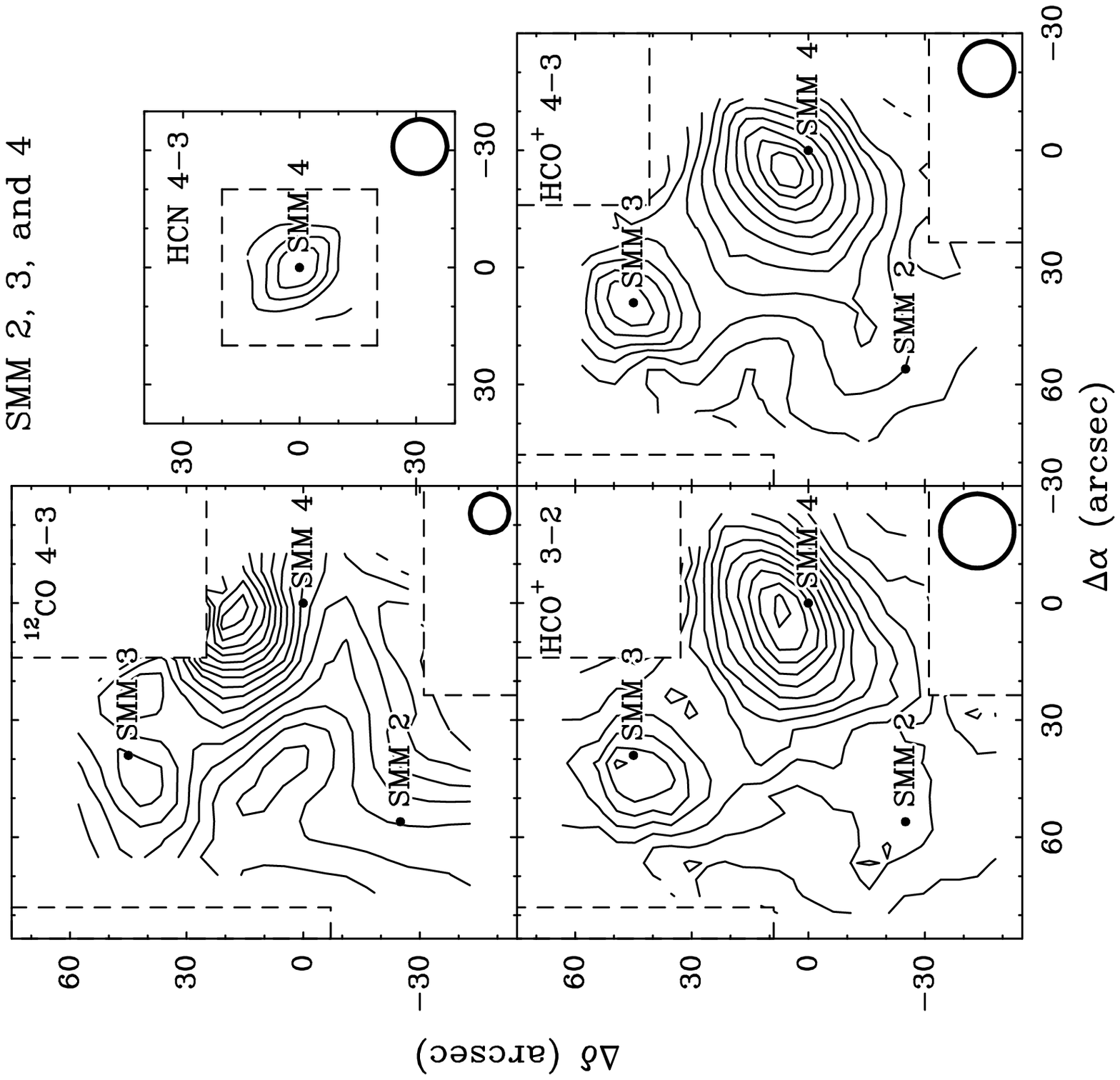]{({\it a\/}) Maps of integrated
intensity observed with the JCMT toward SMM~1. Sampling is indicated
by the small dots, except for HCO$^+$ 3--2, which is fully
sampled. The dashed lines shows the extent of the mapped region. The
beam sizes are indicated in each panel. Contours are drawn at
approximately 3$\sigma$ intervals of 20 ($^{12}$CO), 4 (HCO$^+$ 4--3),
2.4 (HCO$^+$ 3--2, HCN 4--3), 0.9 (H$^{13}$CO$^+$), and 0.5 (H$_2$CO,
SO) K~km~s$^{-1}$. The thick contour indicates the 50\% of maximum
intensity level. ({\it b\/}) Same, for SMM~2, 3, and 4 together. The
sampling, identical to that for SMM~1, is not indicated for clarity.
Contour levels are 15 K~km~s$^{-1}$ for $^{12}$CO and 3 K~km~s$^{-1}$
for HCO$^+$ 3--2, 4--3 and HCN.
\label{f7}}



\newpage

\begin{deluxetable}{lrrrrr}
\tablecolumns{6}
\tablewidth{0pt}
\scriptsize
\tablecaption{Source sample\label{t1}}
\tablehead{
 &
\colhead{$\alpha$ (1950.0)\tablenotemark{a}} & 
\colhead{$\delta$ (1950.0)\tablenotemark{a}} & 
\colhead{$L_{\rm bol}$\tablenotemark{b}} & \colhead{$F_\nu$ (1.1 mm)} &
\colhead{$M_\star$\tablenotemark{c}} \nl
\colhead{Source} &
\colhead{(hh mm ss)} & \colhead{($^\circ$ $'$ $''$)} &
\colhead{($L_{\sun}$)} & \colhead{(Jy)} &
\colhead{($M_\odot$)}}
\startdata
SMM~1 = FIRS~1 & 18 27 17.3 & +01 13 16 & 77 & $3.47 \pm 0.1$ & 0.7--3.9 \nl
SMM~2 & 18 27 28.0\tablenotemark{d} & +01 10 45\tablenotemark{d} & 
10 & $0.6 \pm 0.1$ & 0.1--2.1 \nl
SMM~3 & 18 27 26.9 & +01 11 55 & 13 & $1.11 \pm 0.1$ & 0.1--2.2 \nl
SMM~4 & 18 27 24.3 & +01 11 10 & 15 & $1.47 \pm 0.1$ & 0.1--2.3 \nl
\enddata
\tablenotetext{a}{Best-fit positions to continuum interferometric data (see \S 4.1).} 
\tablenotetext{b}{Scaled to a distance of 400 pc.} 
\tablenotetext{c}{Constraints on stellar mass derived from bolometric
luminosity. Lower limit: $L_{\rm bol}$ due to accretion at $\dot M =
10^{-5}$ $M_\odot$~yr$^{-1}$ onto star with 3 $R_\odot$ radius. Upper 
limit: main-sequence mass of star with stellar luminosity equal to 
$L_{\rm bol}$.} 
\tablenotetext{d}{No continuum emission detected in interferometer beam. 
Submillimeter position of Casali et~al.\ (1993) is given.} 
\tablerefs{Casali et~al.\ 1993 ($F_\nu$ 1.1 mm); Hurt \& Barsony 1996 
($L_{\rm bol}$).} 
\end{deluxetable}

\begin{deluxetable}{lll}
\tablecolumns{3}
\tablewidth{0pt}
\scriptsize
\tablecaption{Overview of observations\label{t2}}
\tablehead{
\colhead{Date} & \colhead{Instrument} & \colhead{Observation}}
\startdata
\cutinhead{Interferometer data}
1994 Oct--Dec, 1996 Feb--May, 1997 Feb--Apr & OVRO
& $F_\nu(\lambda=3.4$ mm) \nl
1997 Apr & OVRO & $F_\nu(\lambda=3.2$ mm)\tablenotemark{a} \nl
1995 Feb, May, 1997 Feb--Mar & OVRO
& $F_\nu(\lambda=2.7$ mm) \nl
1997 Feb--Mar & OVRO & $F_\nu(\lambda=1.3$ mm)\tablenotemark{b} \nl
1997 Apr & OVRO & $F_\nu(\lambda=1.4$ mm)\tablenotemark{a} \nl
1995 Feb, May, 1997 Feb--Apr & OVRO & $^{13}$CO 1--0; C$^{18}$O 1--0 \nl
1994 Oct--Dec & OVRO & HCO$^+$ 1--0; H$^{13}$CO$^+$ 1--0; SiO 2--1 \nl
1996 Feb--May, 1997 Feb--Apr & OVRO & HCN 1--0; H$^{13}$CN 1--0;
C$_3$H$_2$ $4_{32}$--$4_{23}$; SO $2_2$--$1_1$ \nl
1997 Apr & OVRO & N$_2$H$^+$ 1--0\tablenotemark{a}\ ; 
C$^{34}$S 2--1\tablenotemark{a} \nl
\cutinhead{Single-dish data}
1995 May & CSO & $^{12}$CO 6--5\tablenotemark{d}\ ; 
$^{13}$CO 6--5\tablenotemark{d} \nl
1995 Oct & CSO & H$^{13}$CO$^+$ 4--3; SO $8_8$--$7_7$ \nl
1995 Mar, Jun & JCMT & $^{12}$CO 4--3\tablenotemark{e}\ ; 
$^{13}$CO 2--1, 3--2;
 C$^{18}$O 2--1; C$^{17}$O 3--2; SiO 6--5; \nl
 & & HCO$^+$ 3--2\tablenotemark{e}\ , 4--3\tablenotemark{e}\ ; 
H$^{13}$CO$^+$ 3--2\tablenotemark{f}\ ; HC$_3$N
24--23\tablenotemark{f}\ ; 
SO $5_5$--$4_4$\tablenotemark{f,g}\ \ \ ; \nl
 & & CN $3{5\over 2}{3\over 2}$--$2{3\over 2}{1\over 2}$,
$3{5\over 2}{5\over 2}$--$2{3\over 2}{3\over 2}$,
$3{5\over 2}{7\over 2}$--$2{3\over 2}{5\over 2}$,
$3{5\over 2}{3\over 2}$--$2{3\over 2}{3\over 2}$,
$3{5\over 2}{5\over 2}$--$2{3\over 2}{5\over 2}$; \nl
 & & H$_2$CO $7_{17}$--$6_{16}$, $5_{15}$--$4_{14}$,
$5_{33}$--$4_{32}$\tablenotemark{c}\ ,
$5_{32}$--$4_{31}$\tablenotemark{c}\ ,
$3_{03}$--$2_{02}$\tablenotemark{f}\ , 
$3_{22}$--$2_{21}$\tablenotemark{f,g}\ \ \ ; \nl
 & & C$_3$H$_2$ $5_{24}$--$4_{13}$\tablenotemark{f}\ , $5_{32}$--$4_{41}$;
CH$_3$OH $4_2$--$3_1$ E\tablenotemark{f}\ ; 
C~I $^3{\rm P}_1$--$^3{\rm P}_0$ \nl
1996 Aug & JCMT & HCN 4--3; H$^{13}$CN 4--3 \nl
\enddata 
\tablenotetext{a}{SMM~1 and SMM~4 only.} 
\tablenotetext{b}{SMM~2 and SMM~3 only.} 
\tablenotetext{c}{SMM~1 only.} 
\tablenotetext{d}{Five-point map obtained.} 
\tablenotetext{e}{Map obtained.} 
\tablenotetext{f}{Map obtained of SMM~1 only.} 
\tablenotetext{g}{H$_2$CO $3_{22}$--$2_{21}$ and SO $5_5$--$4_4$ 
are partially blended, in upper and lower sideband respectively.} 
\end{deluxetable}

\begin{deluxetable}{lrrrr}
\tablecolumns{5}
\tablewidth{0pt}
\scriptsize
\tablecaption{Millimeter-continuum data\label{t3}}
\tablehead{
 & \colhead{$\lambda$} & \colhead{beam} & \colhead{$I$(max)} &
\colhead{$F_\nu$(total)} \nl
\colhead{Source} & \colhead{(mm)} & \colhead{($''$)} &
\colhead{(Jy~beam$^{-1}$)} & \colhead{(Jy)}}
\startdata
SMM~1 & 3.4 & $5.8\times 4.9$ & 0.126 & 0.202 \nl
& 3.2 & $2.4\times 1.5$ & 0.122 & 0.204 \nl
& 2.7 & $3.2\times 2.7$ & 0.183 & 0.414 \nl
& 1.4 & $1.1\times 0.7$ & 0.615 & 2.650 \nl
SMM~2 & 3.4 & $5.3\times 5.0$ & $<0.005$ & \nodata \nl
& 2.7 & $4.2\times 3.6$ & $<0.008$ & \nodata \nl
& 1.4 & $2.4\times 1.5$ & $<0.11$ & \nodata \nl
SMM~3 & 3.4 & $5.2\times 4.4$ & 0.043 & 0.052 \nl
& 2.7 & $4.3\times 3.6$ & 0.090 & 0.104 \nl
& 1.4 & $2.4\times 1.5$ & 0.651 & 0.900 \nl
SMM~4 & 3.4 & $5.3\times 4.6$ & 0.060 & 0.075 \nl
& 3.2 & $2.6\times 1.5$ & 0.068 & 0.097 \nl
& 2.7 & $3.2\times 2.7$ & 0.092 & 0.143 \nl
& 1.4 & $1.1\times 0.7$ & 0.462 & 1.108 \nl
\enddata
\end{deluxetable}

\begin{deluxetable}{lrrr}
\tablecolumns{4}
\tablewidth{0pt}
\scriptsize
\tablecaption{Envelope model and best-fit parameters\label{t4}}
\tablehead{
 & \colhead{SMM 1} & \colhead{SMM 3} & \colhead{SMM 4}}
\startdata
$R_{\rm in}$ (AU)\tablenotemark{a} & 100  & 100  & 100  \nl
$R_{\rm out}$ (AU)\tablenotemark{a} & 8000  & 8000  & 8000  \nl
\cutinhead{Density: $\rho(r) = \rho_0 (r/1000\,{\rm AU})^{-p}$}
$p$ & $-2.0 \pm 0.5$ & $-2.0\pm 0.5$ & $-2.0\pm 0.5$ \nl
$M_{\rm env}$($M_\odot$) & 8.7 & 3.0 & 5.3 \nl
\cutinhead{Dust temperature: $T_{\rm dust} = T_0 (r/1000\,{\rm AU})^{-0.4}$}
$T_0$ & 27 K & 24 K & 20 K \nl
\cutinhead{Point-source flux\tablenotemark{b}}
$F_\nu$(2.7 mm) (Jy) & 0.13 & 0.05 & 0.07 \nl
\enddata
\tablenotetext{a}{Fixed parameter.}
\tablenotetext{b}{Additional unresolved component to fit flux on long
baselines. See text.}
\end{deluxetable}

\begin{deluxetable}{lll}
\tablecolumns{3}
\tablewidth{0pt}
\scriptsize
\tablecaption{Overview of regions traced by the various observations
presented in this paper\label{t5}}
\tablehead{
\colhead{Component} & \colhead{Interferometer} & \colhead{Single dish}}
\startdata
Bulk of the (cold) envelope &
  Continuum on short spacings &
  $^{13}$CO 2--1, 3--2; C$^{18}$O 2--1; C$^{17}$O 3--2 \nl
 & & HCO$^+$ 3--2, 4--3 (line center); H$^{13}$CO$^+$ 3--2, 4--3 \nl
 & & HCN 4--3 (line center); H$^{13}$CN 4--3 \nl
 & & H$_2$CO $3_{03}$--$2_{02}$, $5_{15}$--$4_{14}$, $7_{17}$--$6_{16}$ \nl
 & & C$_3$H$_2$ $5_{32}$--$4_{41}$; HC$_3$N 24--23; 
     CN $3{5\over2}$--$2{3\over2}$ \nl
 & & SiO 6--5\tablenotemark{a}; SO $5_5$--$4_4$, $8_8$--$7_7$\tablenotemark{a} \nl
Inner regions of the envelope & Continuum on intermediate spacings & \nodata
 \nl
 & $^{13}$CO, C$^{18}$O 1--0 & \nl
 & H$^{13}$CO$^+$, H$^{13}$CN, N$_2$H$^+$ 1--0 & \nl
Additional warm material\tablenotemark{b} & Continuum on longest spacings &
  $^{12}$CO 4--3, 6--5; $^{13}$CO 6--5; H$_2$CO $3_{22}$--$2_{21}$ \nl
Bipolar outflow & SiO 2--1; SO $2_2$--$1_1$ & 
  $^{12}$CO 4--3, 6--5 (wings); $^{13}$CO 6-5 (wings)  \nl
 & & HCN 4--3 (wings) \nl
Walls of the outflow cavity & HCO$^+$ 1--0; HCN 1--0 & \nodata \nl
Surrounding quiescent cloud &
  C$_3$H$_2$ $4_{32}$--$4_{23}$ (clumps) & 
  [C~I] $^3$P$_1$--$^3$P$_0$ (surface) \nl 
\enddata
\tablenotetext{a}{For the observations of SMM~1 do the single-dish SiO
and SO observations trace the envelope only, because the beam did not
cover the SiO and SO peak due to the outflow seen in the OVRO images.}
\tablenotetext{b}{Associated with the innermost few hundred AU of the
envelopes, where the temperature exceeds the adopted $T_{\rm kin}\propto 
r^{-0.4}$ relation (see text).}
\end{deluxetable}

\begin{deluxetable}{llrrrrr}
\tablecolumns{7}
\tablewidth{0pt}
\scriptsize
\tablecaption{OVRO integrated intensities and line opacities\label{t6}}
\tablehead{
 & & \multicolumn{2}{c}{$5''\times 5''$ area} & & \multicolumn{2}{c}{$20''\times 20''$ area}\nl
\cline{3-4} \cline{6-7}\nl
 & & 
\colhead{$\int T_{\rm b}dV$} & &&
\colhead{$\int T_{\rm b}dV$} \nl
\colhead{Source} & \colhead{Line} & \colhead{(K~km~s$^{-1}$)} &
\colhead{$\bar\tau$} &&
\colhead{(K~km~s$^{-1}$)} &
\colhead{$\bar\tau$}}
\startdata
SMM~1 &
$^{13}$CO 1--0 & $27.1\pm 0.3$ & 2.1 &&
$2.49\pm 0.08$ & 6.5 \nl
& C$^{18}$O 1--0 & $7.4\pm 0.1$ & 0.3 &&
$1.38\pm 0.03$ & 0.8 \nl
& C$^{34}$S 2--1 & $<3.9$ & \nodata &&
$<1.0$ & \nodata \nl
& C$_3$H$_2$ $4_{32}$--$4_{23}$ & $2.0\pm 0.1$ & \nodata &&
$0.34\pm 0.02$ & \nodata \nl
& HCO$^+$ 1--0 & $48.4\pm 0.1$ & 19.5 &&
$14.0\pm 0.03$ & 7.7 \nl
& H$^{13}$CO$^+$ 1--0 & $12.5\pm 0.2$ & 0.3 &&
$1.56\pm 0.06$ & 0.1 \nl
& HCN 1--0 & $22.0\pm 0.1$ & 15.5 &&
$4.96\pm 0.03$ & 5.3 \nl
& H$^{13}$CN 1--0 & $4.7\pm 0.1$ & 0.2 &&
$0.39\pm 0.03$ & 0.1 \nl
& N$_2$H$^+$ 1--0 & $1.4\pm 0.1$ & \nodata &&
$3.79\pm 0.22$ & $2\pm 1$ \nl
& SiO 2--1 & $3.9\pm 0.1$ & \nodata &&
$1.62\pm 0.03$ & \nodata \nl
& SO $2_2$--$1_1$ & $2.0\pm 0.1$ & \nodata &&
$0.43\pm 0.01$ & \nodata \nl
SMM~2
& $^{13}$CO 1--0 & $<0.5$ & \nodata &&
$0.25\pm 0.05$ & $<5.5$ \nl
& C$^{18}$O 1--0 & $<0.3$ & \nodata &&
$<0.1$ & $<0.7$ \nl
& C$_3$H$_2$ $4_{32}$--$4_{23}$ & $<0.4$ & \nodata &&
$<0.10$ & \nodata \nl
& HCO$^+$ 1--0 & $<0.5$ & \nodata &&
$0.64\pm 0.04$ & \nodata \nl
& H$^{13}$CO$^+$ 1--0 & $1.2\pm 0.1$ & \nodata &&
$1.23\pm 0.03$ & \nodata \nl
& HCN 1--0 & $1.2\pm 0.2$ & \nodata &&
$1.53\pm 0.04$ & $5\pm 4$ \nl
& H$^{13}$CN 1--0 & $<0.8$ & \nodata &&
$<0.2$ & $<0.1$ \nl
& SiO 2--1 & $<1.3$ & \nodata &&
$<0.3$ & \nodata \nl
& SO $2_2$--$1_1$ & $<0.5$ & \nodata &&
$<0.1$ & \nodata \nl
SMM~3
& $^{13}$CO 1--0 & $11.8\pm 0.2$ & $<5.5$ &&
$3.37\pm 0.05$ & $<0.2$ \nl
& C$^{18}$O 1--0 & $<0.45$ & $<0.7$ &&
$<0.15$ & $<0.1$ \nl
& C$_3$H$_2$ $4_{32}$--$4_{23}$ & $<0.33$ & \nodata &&
$<0.12$ & \nodata \nl
& HCO$^+$ 1--0 & $9.3\pm 0.2$ & $<2.4$ &&
$4.07\pm 0.05$ & $<1.1$ \nl
& H$^{13}$CO$^+$ 1--0 & $<0.36$ & $<0.04$ &&
$<0.1$ & $<0.1$ \nl
& HCN 1--0 & $33.2\pm 0.2$ & $<0.5$ &&
$12.8\pm 0.06$ & 2.8 \nl
& H$^{13}$CN 1--0 & $<0.9$ & $<0.1$ &&
$0.7\pm 0.1$ & 3.5 \nl
& SiO 2--1 & $<1.8$ & \nodata &&
$<0.51$ & \nodata \nl
& SO $2_2$--$1_1$ & $<0.4$ & \nodata &&
$<0.1$ & \nodata \nl
SMM~4
& $^{13}$CO 1--0 & $<0.75$ & \nodata &&
$<0.21$ & \nodata \nl
& C$^{18}$O 1--0 & $4.4\pm 0.1$ & \nodata &&
$0.82\pm 0.02$ & \nodata \nl
& C$^{34}$S 2--1 & $<5.0$ & \nodata &&
$<1.5$ & \nodata \nl
& C$_3$H$_2$ $4_{32}$--$4_{23}$ & $0.5\pm 0.1$ & \nodata &&
$<0.06$ & \nodata \nl
& HCO$^+$ 1--0 & $40.8\pm 0.2$ & 9.3 &&
$18.0\pm 0.06$ & 4.2 \nl
& H$^{13}$CO$^+$ 1--0 & $5.4\pm 0.1$ & 0.1 &&
$1.13\pm 0.04$ & $<0.1$ \nl
& HCN 1--0 & $31.4\pm 0.2$ & $<0.5$ &&
$13.3\pm 0.04$ & $<0.2$ \nl
& H$^{13}$CN 1--0 & $<0.6$ & $<0.1$ &&
$<0.15$ & $<0.1$ \nl
& N$_2$H$^+$ 1--0 & $7.8\pm 1.2$ & $<2.0$ &&
$<1.0$ & \nodata \nl
& SiO 2--1 & $<1.2$ & \nodata &&
$0.28\pm 0.11$ & \nodata \nl
& SO $2_2$--$1_1$ & $<0.3$ & \nodata &&
$<0.06$ & \nodata \nl
\enddata
\end{deluxetable}

\begin{deluxetable}{lrrrrrrrrrrr}
\tablecolumns{12}
\tablewidth{0pt}
\scriptsize
\tablecaption{Single-dish integrated line intensities and opacities\label{t7}}
\tablehead{
 & \multicolumn{2}{c}{SMM~1} && \multicolumn{2}{c}{SMM~2} &&
 \multicolumn{2}{c}{SMM~3} && \multicolumn{2}{c}{SMM~4} \nl
\cline{2-3} \cline{5-6} \cline{8-9} \cline{11-12}\nl
 & 
\colhead{$\int T_{\rm mb}dV$\tablenotemark{a}} & 
\colhead{$\bar\tau$\tablenotemark{b}} &&
\colhead{$\int T_{\rm mb}dV$\tablenotemark{a}} & 
\colhead{$\bar\tau$\tablenotemark{b}} &&
\colhead{$\int T_{\rm mb}dV$\tablenotemark{a}} & 
\colhead{$\bar\tau$\tablenotemark{b}} &&
\colhead{$\int T_{\rm mb}dV$\tablenotemark{a}} & 
\colhead{$\bar\tau$\tablenotemark{b}} \nl
\colhead{Line} & 
\colhead{(K~km~s$^{-1}$)} & &&
\colhead{(K~km~s$^{-1}$)} & &&
\colhead{(K~km~s$^{-1}$)} & &&
\colhead{(K~km~s$^{-1}$)} & 
}
\startdata
 $^{12}$CO 4--3 
& $232.2 \pm 2.2$ & \nodata
&& $120.6 \pm 1.7$ & \nodata
&& $171.3 \pm 1.8$ & \nodata
&& $148.5 \pm 1.6$ & \nodata \nl
 \hphantom{$^{12}$CO} 6--5 
& $221.9 \pm 2.7$ & 13.5 
&& $60.1 \pm 1.8$ & 6.4 
&& $165.7 \pm 2.0$ & 5.2
&& $134.5 \pm 1.4$ & 4.6 \nl
 $^{13}$CO 2--1 
& $27.0 \pm 0.2$ & 3.4
&& $40.8 \pm 0.3$ & 5.5
&& $43.1 \pm 0.3$ & 0.9
&& $42.8 \pm 0.3$ & 2.0 \nl
 \hphantom{$^{13}$CO} 3--2 
& $35.8 \pm 0.6$ & 7.3 
&& $39.9 \pm 0.8$ & 5.5 
&& $50.1 \pm 0.6$ & 2.4 
&& $45.7 \pm 0.7$ & 4.0 \nl
 \hphantom{$^{13}$CO} 6--5 
& $41.4 \pm 1.0$ & 0.2
&& $5.6 \pm 0.7$ & 0.1 
&& $12.8 \pm 0.6$ & 0.1
&& $9.2 \pm 0.5$ & 0.1 \nl
 C$^{18}$O 2--1 
& $9.5 \pm 0.2$ & 0.4 
&& $13.0 \pm 0.3$ & 0.3 
&& $7.7 \pm 0.3$ & 0.1
&& $11.0 \pm 0.3$ & 0.3 \nl
 C$^{17}$O 3--2 
& $5.7 \pm 0.3$ & 0.2
&& $4.9 \pm 0.2$ & 0.1
&& $3.0 \pm 0.2$ & 0.1
&& $4.2 \pm 0.2$ & 0.1 \nl
 HCO$^+$ 3--2 
& $33.3 \pm 0.4$ & 12.4
&& $13.1 \pm 0.2$ & 9.1
&& $21.8 \pm 0.3$ & $5\pm 3$
&& $36.0 \pm 0.5$ & 5.5 \nl
 \hphantom{HCO$^+$} 4--3 
& $48.1 \pm 0.7$ & 4.3
&& $8.8 \pm 0.5$ & $<3.1$ 
&& $20.4 \pm 0.8$ & $<0.2$ 
&& $25.0 \pm 0.6$ & 3.0 \nl
 H$^{13}$CO$^+$ 3--2 
& $5.8 \pm 0.2$ & 0.2
&& $1.7 \pm 0.1$ & 0.1 
&& $1.6 \pm 0.8$ & 0.1
&& $2.9 \pm 0.1$ & 0.1 \nl
 \hphantom{H$^{13}$CO$^+$} 4--3 
& $3.1 \pm 0.1$ & 0.1
&& $<0.4$ & $<0.1$ 
&& $0.2 \pm 0.1$ & $<0.1$
&& $1.2 \pm 0.1$ & 0.1 \nl
 HCN 4--3 
& $17.5 \pm 0.4$ & 12.3
&& $1.4 \pm 0.3$ & \nodata
&& $10.1 \pm 0.3$ & $<0.2$ 
&& $14.0 \pm 0.4$ & $<2.1$ \nl
 H$^{13}$CN 4--3 
& $3.0\pm 0.2$ & 0.2 
&& \nodata & \nodata
&& $<0.8$ & $<0.1$ 
&& $<0.5$ & $<0.1$ \nl
 H$_2$CO $3_{03}$--$2_{02}$ 
& $4.6 \pm 0.1$ & \nodata
&& $2.7 \pm 0.1$ & \nodata
&& $4.9 \pm 0.1$ & \nodata
&& $11.0 \pm 0.1$ & \nodata \nl
 \hphantom{H$_2$CO} $3_{22}$--$2_{21}$\tablenotemark{d} 
& $1.8 \pm 0.1$ & \nodata
&& $0.4 \pm 0.1$ & \nodata
&& $1.7 \pm 0.1$ & \nodata
&& $2.93 \pm 0.1$ & \nodata \nl
 \hphantom{H$_2$CO} $5_{15}$--$4_{14}$ 
& $8.0 \pm 0.3$ & \nodata
&& $0.7 \pm 0.2$ & \nodata
&& $7.2 \pm 0.3$ & \nodata
&& $6.4 \pm 0.3$ & \nodata \nl
 \hphantom{H$_2$CO} $5_{32}$--$4_{31}$ 
& $3.3 \pm 0.3$ & \nodata
&& \nodata & \nodata
&& \nodata & \nodata
&& \nodata & \nodata \nl
 \hphantom{H$_2$CO} $5_{33}$--$4_{32}$ 
& $2.8 \pm 0.3$ & \nodata
&& \nodata & \nodata
&& \nodata & \nodata
&& \nodata & \nodata \nl
 \hphantom{H$_2$CO} $7_{17}$--$6_{16}$ 
& $6.7 \pm 0.6$ & \nodata
&& $<1.5$ & \nodata
&& $<1.3$ & \nodata
&& $<1.6$ & \nodata \nl
 C~I $^3{\rm P}_1$--$^3{\rm P}_0$ 
& $18.0 \pm 0.7$ & \nodata
&& $26.7 \pm 1.0$ & \nodata
&& $30.3 \pm 1.0$ & \nodata
&& $29.8 \pm 0.9$ & \nodata \nl
 C$_3$H$_2$ $5_{24}$--$4_{13}$ 
& $0.7 \pm 0.1$ & \nodata
&& $<0.3$ & \nodata
&& $<0.2$ & \nodata
&& $<0.2$ & \nodata \nl 
 \hphantom{C$_3$H$_2$} $5_{32}$--$4_{41}$ 
& $1.0 \pm 0.1$ & \nodata
&& $<0.3$ & \nodata
&& $<0.2$ & \nodata
&& $<0.3$ & \nodata \nl
 CH$_3$OH $4_2$--$3_1$ E 
& $1.4 \pm 0.1$ & \nodata 
&& $<0.3$ & \nodata
&& $0.6 \pm 0.1$ & \nodata
&& $2.1 \pm 0.1$ & \nodata \nl
 HC$_3$N 24--23 
& $0.9 \pm 0.1$ & \nodata
&& $<0.5$ & \nodata
&& $<0.2$ & \nodata
&& $<0.2$ & \nodata \nl
 CN $3{5\over 2}{3\over 2}$--$2{3\over 2}{1\over 2}$, ${5\over 2}$--${3\over 2}$\tablenotemark{c} 
& $6.6 \pm 0.3$ & \nodata
&& $1.4 \pm 0.2$ & \nodata
&& $0.7 \pm 0.2$ & \nodata
&& $1.4 \pm 0.2$ & \nodata \nl
 \hphantom{CN} $3{5\over 2}{3\over 2}$--$2{3\over 2}{3\over 2}$ 
& $0.8 \pm 0.2$ & \nodata
&& $<0.6$ & \nodata
&& $<0.6$ & \nodata
&& $<0.5$ & \nodata \nl
 \hphantom{CN} $3{5\over 2}{5\over 2}$--$2{3\over 2}{5\over 2}$ 
& $<0.8$ & \nodata
&& $<0.6$ & \nodata
&& $<0.6$ & \nodata
&& $<0.5$ & \nodata \nl
 \hphantom{CN} $3{5\over 2}{7\over 2}$--$2{3\over 2}{5\over 2}$\tablenotemark{c} 
& $6.0 \pm 0.2$ & \nodata
&& $<0.6$ & \nodata 
&& $0.6 \pm 0.2$ & \nodata
&& $0.8 \pm 0.2$ & \nodata \nl
 SO $5_5$--$4_4$\tablenotemark{d} 
& $1.4 \pm 0.1$ & \nodata
&& $<0.3$ & \nodata
&& $<0.1$ & \nodata
&& $<0.2$ & \nodata \nl
 \hphantom{SO} $8_8$--$7_7$ 
& $0.3 \pm 0.1$ & \nodata
&& $<0.3$ & \nodata
&& $0.3 \pm 0.1$ & \nodata
&& $<0.3$ & \nodata \nl
 SiO 6--5
& $0.5\pm 0.1$  & \nodata
&& $<0.3$ & \nodata
&& $<0.2$ & \nodata
&& $<0.3$ & \nodata \nl
\enddata
\tablenotetext{a}{Upper limits are 3$\sigma$.} 
\tablenotetext{b}{Using abundance ratios ${\rm [^{12}C]:[^{13}C]=65:1}$
and ${\rm [^{16}O]:[^{18}O]:[^{17}O]=2695:5:1}$.} 
\tablenotetext{c}{CN $3{5\over 2}{3\over 2}$--$2{3\over 2}{1\over 2}$ and 
$3{5\over 2}{5\over 2}$--$2{3\over 2}{3\over 2}$ are partially blended with
$3{5\over 2}{7\over 2}$--$2{3\over 2}{5\over 2}$. Assuming equal contributions to
integrated intensity.} 
\tablenotetext{d}{H$_2$CO $3_{22}$--$2_{21}$ and SO $5_5$--$4_4$ are partially
blended. Assuming equal contributions for SMM~1, and no contribution from SO for
SMM~2, 3, and 4 (cf.\ spectra in Fig.~6).} 
\end{deluxetable}

\begin{deluxetable}{lrrrrrrrr}
\tablecolumns{9}
\tablewidth{0pt}
\scriptsize
\tablecaption{Derived molecular abundances\label{t8}}
\tablehead{
 & \multicolumn{3}{c}{Envelope} && 
   \multicolumn{3}{c}{Warm gas\tablenotemark{a}} \nl
\cline{2-4} \cline{6-8} \nl
\colhead{Species} & 
   \colhead{SMM~1} & \colhead{SMM~3} &\colhead{SMM~4} &&
   \colhead{SMM~1} & \colhead{SMM~3} &\colhead{SMM~4} & 
   \colhead{IRAS 16293$-$2422\tablenotemark{b}}
}
\startdata
$^{12}$CO\tablenotemark{c} & 
   $\equiv 1(-4)$ & $\equiv 1(-4)$ & $\equiv 1(-4)$ &&
   $\equiv 1(-4)$ & $\equiv 1(-4)$ & $\equiv 1(-4)$ & \nodata \nl
HCO$^+$ & $1(-9)$ & $1(-9)$ & $1(-9)$ && $2(-8)$ & $2(-8)$ & $1(-7)$ & 
   $2(-9)$ \nl
HCN & $2(-9)$ & $2(-9)$ & $2(-9)$ && $5(-8)$ & $1(-7)$ & $8(-8)$ &
   $2(-9)$ \nl
H$_2$CO & $8(-10)$ & $2(-9)$ & $2(-9)$ && $9(-9)$ & $2(-8)$ & $1(-7)$ &
   $7(-10)$ \nl
C$_3$H$_2$ & $2(-10)$ & $<3(-10)$ & $<3(-10)$ && $3(-9)$ & $<3(-9)$ & 
   $<7(-9$) & $4(-11)$ \nl
CN & $5(-9)$ & $2(-10)$ & $2(-10)$ && $3(-8)$ & $1(-8)$ & $1(-7)$ &
   $1(-10)$ \nl
HC$_3$N & $2(-10)$ & $<2(-10)$ & $<2(-10)$ && $9(-10)$ & $<4(-9)$ & $<6(-9)$ & 
   $3(-11)$ \nl
SiO & $1(-11)$ & $<2(-11)$ & $<2(-11)$ && $1(-10)$ & $<4(-10)$ & $<1(-9)$ & 
   $1(-10)$ \nl
SO & $2(-10)$ & $<1(-10)$ & $<2(-10)$ && $2(-9)$ & $<2(-9)$ & $<7(-9)$ &
   $4(-9)$ \nl
C~I\tablenotemark{d} & 
   $9(-6)$ & $5(-5)$ & $3(-5)$ && $4(-5)$ & $3(-4)$ & $8(-4)$ & \nodata \nl
\enddata
\tablenotetext{a}{Abundances derived under the assumption that all emission
except CO originates in the additional column of $\sim 100$ K gas. See 
text.}
\tablenotetext{b}{Average abundances in a $20''$ beam toward the
class~0 YSO IRAS 16293$-$2422, for comparison (van Dishoeck et~al.\ 1995).}
\tablenotetext{c}{Abundance of CO is fixed, except for depletion. See
text.}  
\tablenotetext{d}{Derived abundance of C~I is highly uncertain,
because the low-density surface of the entire Serpens Molecular Cloud
is likely to contribute significantly to the emission in the
$^3$P$_1$--$^3$P$_0$ line.}
\end{deluxetable}



\null\newpage
\input psfig

\rightline{\underline{\tt Fig. 1}}
\centerline{\psfig{figure=fig1.ps,rheight=10truecm,height=22truecm}}
\vfill\eject
\rightline{\underline{\tt Fig. 2}}
\centerline{\psfig{figure=fig2.ps,rheight=10truecm,height=22truecm}}
\vfill\eject
\rightline{\underline{\tt Fig. 3}}
\centerline{\psfig{figure=fig3.ps,rheight=10truecm,height=22truecm}}
\vfill\eject
\rightline{\underline{\tt Fig. 4a}}
\centerline{\psfig{figure=fig4a.ps,rheight=10truecm,height=22truecm}}
\vfill\eject
\rightline{\underline{\tt Fig. 4b}}
\centerline{\psfig{figure=fig4b.ps,rheight=10truecm,height=22truecm}}
\vfill\eject
\rightline{\underline{\tt Fig. 5}}
\centerline{\psfig{figure=fig5.ps,rheight=10truecm,height=22truecm}}
\vfill\eject
\rightline{\underline{\tt Fig. 6a}}
\centerline{\psfig{figure=fig6a.ps,rheight=10truecm,height=22truecm}}
\vfill\eject
\rightline{\underline{\tt Fig. 6b}}
\centerline{\psfig{figure=fig6b.ps,rheight=10truecm,height=22truecm}}
\vfill\eject
\rightline{\underline{\tt Fig. 6c}}
\centerline{\psfig{figure=fig6c.ps,rheight=10truecm,height=22truecm}}
\vfill\eject
\rightline{\underline{\tt Fig. 6d}}
\centerline{\psfig{figure=fig6d.ps,rheight=10truecm,height=22truecm}}
\vfill\eject
\rightline{\underline{\tt Fig. 7a}}
\centerline{\psfig{figure=fig7a.ps,rheight=10truecm,height=22truecm}}
\vfill\eject
\rightline{\underline{\tt Fig. 7b}}
\centerline{\psfig{figure=fig7b.ps,rheight=10truecm,height=22truecm}}
\vfill\eject

\end{document}